\documentclass[numberedappendix, onecolumn]{emulateapj} 
\usepackage{rotating} 
\usepackage{amsmath,amssymb}

\begin{document}

\title{Scattering Polarization of the Ca {\sc ii} IR Triplet for Probing the Quiet Solar Chromosphere}

\author{R. Manso Sainz\altaffilmark{1,2} and J. Trujillo Bueno\altaffilmark{1,2,3}} \altaffiltext{1}{Instituto de Astrof\'{\i}sica de Canarias, 38205, La Laguna, Tenerife, Spain}\altaffiltext{2}{Departamento de Astrof\'\i sica, Facultad de F\'\i sica, Universidad de La Laguna, Tenerife, Spain}\altaffiltext{3}{Consejo Superior de Investigaciones Cient\'{\i}ficas, Spain} \email{rsainz@iac.es, jtb@iac.es}

\begin{abstract}
The chromosphere of the quiet Sun is a very important stellar atmospheric region whose thermal and magnetic structure we need to decipher for unlocking new discoveries in solar and stellar physics. To this end, we need to identify and exploit observables sensitive to weak magnetic fields ($B{\lesssim}100$ G) and to the presence of cool and hot gas in the bulk of the solar chromosphere. Here we report on an investigation of the Hanle effect in two semi-empirical models of the quiet solar atmosphere with different chromospheric thermal structures. Our study reveals that the linear polarization profiles produced by scattering in the Ca {\sc ii} IR triplet have thermal and magnetic sensitivities potentially of great diagnostic value. The linear polarization in the 8498 \AA\ line shows a strong sensitivity to inclined magnetic fields with strengths between 0.001 and 10 G, while the emergent linear polarization in the 8542~\AA\ and 8662~\AA\ lines is mainly sensitive to magnetic fields with strengths between 0.001 and 0.1 G. The reason for this is that the scattering polarization of the 
8542~\AA\ and 8662~\AA\ lines, unlike the 8498 \AA\ line, is controlled mainly by the Hanle effect in their  (metastable) lower levels. Therefore, in regions with magnetic strengths sensibly larger than 1 G, their Stokes $Q$ and $U$ profiles are sensitive only to the orientation of the magnetic field vector. We also find that for given magnetic field configurations the sign of the $Q/I$ and $U/I$ profiles of the 8542~\AA\ and 8662~\AA\ lines is the same in both atmospheric models, while the sign of the linear polarization profile of the 8498 \AA\ line turns out to be very sensitive to the thermal structure of the lower chromosphere. We suggest that spectropolarimetric observations providing information on the relative scattering polarization amplitudes of the Ca {\sc ii} IR triplet will be very useful to improve our empirical understanding of the thermal and magnetic structure of the quiet chromosphere.
\end{abstract}


%%%%%%%%%%%%%%%%%%%%%%%%%%%%%%%%%%%%%%%%
%%%%%%%%%%%%%%%%%%%%%%%%%%%%%%%%%%%%%%%%
% INTRODUCTION
%%%%%%%%%%%%%%%%%%%%%%%%%%%%%%%%%%%%%%%%
%%%%%%%%%%%%%%%%%%%%%%%%%%%%%%%%%%%%%%%%
\section{Introduction}

The chromosphere of the quiet Sun is one of the most complex stellar atmospheric regions (e.g., Harvey 2006, 2009; Judge 2006, 2009). Lying between the thin photosphere and the extended $10^6$ K corona, it is here where the magnetic field becomes the globally dominating factor ruling the outer solar atmosphere. Measuring the magnetic field vector in the solar chromospheric plasma is, however, notoriously difficult, especially outside sunspots and related active regions. Quantitative information can be obtained through spectropolarimetry, but the measurement and physical interpretation of the weak polarization signatures that the Hanle and Zeeman effects produce in the few chromospheric lines that can be observed from ground-based telescopes is not an easy task (see Trujillo Bueno 2010, for a recent review). It is thus important to develop novel methods of ``measuring" the chromospheric magnetic field, ideally based on the action of the Hanle and Zeeman effects in spectral lines whose line-center intensity images reveal the fine-scale thermodynamic structuring of the solar chromosphere. Here we show that the scattering polarization observed in the IR triplet of Ca {\sc ii} contains valuable information on the thermal and magnetic structure of the ``quiet" chromosphere, from the shock dominated region of the ``lower chromosphere" to the ``upper chromosphere" below the interface region to the $10^6$ K solar corona.

The only way to obtain direct empirical information on the intensity and topology of the magnetic fields of the solar chromosphere is via the measurement and interpretation of polarization signals in chromospheric spectral lines. In regions with high concentrations of magnetic flux, such as in sunspots and plages, the polarization patterns are dominated by the Zeeman effect. Hence, diagnostic techniques based on this effect are quite useful (e.g., Socas-Navarro, Trujillo Bueno \& Ruiz Cobo 2000; Socas-Navarro 2005). However, in the quiet Sun, which covers most of the solar disk at any given time during the solar magnetic activity cycle, the Zeeman splitting between the $\pi$ ($\Delta{M}=M_u-M_l=0$) and ${\sigma}_{b,r}$ ($\Delta{M}={\pm}1$) components ($M$ being the magnetic quantum number) is only a small fraction of the width of the spectral lines formed in the chromosphere. As a result, the transverse Zeeman effect makes a very small or insignificant contribution to the emergent linear polarization amplitudes.
On the other hand, the circular polarization of the Zeeman effect as a diagnostic tool is of limited practical interest for the exploration of the magnetism of the upper chromosphere because the response function of the emergent Stokes $V$ profiles of strong lines like H$\alpha$ and Ca {\sc ii} 8542 \AA\ is significant only in the photosphere and/or lower chromosphere (Socas-Navarro \& Uitenbroek 2004; Uitenbroek 2006). 

There is yet another mechanism producing linear polarization in the spectral lines that can be exploited to diagnose the solar chromosphere. Atoms align due to the optical pumping caused by the incident anisotropic radiation field; i.e., the individual magnetic $M$ sublevels of energy levels with total angular momentum $J>1/2$ are unevenly populated (in such a way that the populations of substates with different values of $|M|$ are unequal) and coherences between them may appear. This, in turn, gives rise to a sizable amount of linear polarization in chromospheric spectral lines, either by selective emission or absorption of polarization components in the spectral line. A remarkable example is provided by the IR triplet of Ca {\sc ii}. Its linear polarization pattern was considered ``enigmatic'' for a while (Stenflo, Keller \& Gandorfer 2000) because it was impossible to achieve even a qualitative fit to the observed relative polarization amplitudes considering that scattering line polarization results only from selective emission processes caused by the population imbalances of the line's upper level. It was later realized that selective absorption of polarization components caused by the presence of populations imbalances in the line's lower level is a very efficient mechanism for producing linear polarization in the spectral lines of a stellar atmosphere (Trujillo Bueno \& Landi Degl'Innocenti 1997; Trujillo Bueno 1999; Trujillo Bueno et al. 2002). In fact, Manso Sainz \& Trujillo Bueno (2003{\natexlab{a}}) demonstrated quantitatively that the physical origin of the linear polarization pattern observed in the Ca {\sc ii} lines at 8542 and 8662 \AA\ is indeed ``zero-field" dichroism, i.e., selective absorption of polarization components caused by the presence of atomic alignment in the metastable lower levels $4^2{\rm D}_{3/2}$ and $4^2{\rm D}_{5/2}$ (see Fig. 1), while that observed in the 8498 \AA\ line is controled by both dichroism and by the selective emission of polarization components that result from the population imbalances between the sublevels of its upper level $^2{\rm P}_{3/2}$. 

The alignment of the atomic $J$-levels and the ensuing emergent linear polarization is modified by the presence of a magnetic field through the Hanle effect, even if the magnetic field turns out to have a complex, tangled topology, with mixed polarities at subresolution scales (e.g., Landi Degl'Innocenti \& Landolfi 2004). In the absence of collisional 
depolarization the magnetic field strength $B$ (in G) that is sufficient to produce a significant change is
\begin{equation}
B\,{\approx}\,{{1.137{\times}10^{-7}}\over{t_{\rm life}g_J}}, 
\end{equation} 
where $t_{\rm life}$ and $g_J$ are, respectively, the lifetime (in seconds) and the Land\'e factor of the $J$ level under consideration. Interestingly enough, since the lifetimes of the metastable lower levels of the Ca {\sc ii} IR triplet are about two orders of magnitude larger than the upper level lifetimes and levels with $J=1/2$ or $J=0$ cannot carry atomic alignment, the linear polarization of the Ca {\sc ii} line at 8662 \AA\ (whose upper level has $J_u=1/2$)
is expected to be sensitive to mG fields only, while the scattering polarization of the 8498 \AA\ and 8542 \AA\ lines (which share the same upper level whose $J_u=3/2$) can in principle be sensitive also to magnetic fields in the gauss range. In the quiet solar chromosphere depolarization of the Ca {\sc ii} atomic levels by elastic collisions with neutral hydrogen atoms is insignificant for the upper levels of the H and K lines, while such collisions have a rather small impact on the atomic polarization of the metastable lower levels (Manso Sainz \& Trujillo Bueno 2003{\natexlab{a}}; Derouich, Trujillo Bueno \& Manso Sainz 2007). Actually, the mere fact that the Ca {\sc ii} 8662 \AA\ line shows a clear $Q/I$ polarization signal (see Stenflo et al. 2000) implies that its lower level (whose $J_l=3/2$) is significantly polarized in the quiet solar chromosphere (Manso Sainz \& Trujillo Bueno 2003{\natexlab{a}}). 

The aim of this paper is to investigate the thermal and magnetic sensitivity of the scattering polarization $Q/I$ and $U/I$ profiles of the Ca {\sc ii} IR triplet. After formulating in \S2 the multilevel problem of scattering polarization and the Hanle effect, \S4 discusses the unmagnetized case in the ``hot" and ``cool" chromospheric models described in \S3. The impact of the Hanle effect on the linear polarization profiles of the Ca {\sc ii} IR triplet is discussed in \S5, showing for each solar atmospheric model detailed results for the cases of disk center (\S5.1) and close to the limb observations (\S5.2). The ensuing information is summarized in \S5.3 in the form of Hanle effect diagrams for each of the spectral lines of the IR triplet. Finally, \S6 summarizes our main conclusions with an outlook to future research.

%%%%%%%%%%%%%%%%%%%%%%%%%%%%%%%%%%%%%%%%
%%%%%%%%%%%%%%%%%%%%%%%%%%%%%%%%%%%%%%%%
% BASIC EQUATIONS
%%%%%%%%%%%%%%%%%%%%%%%%%%%%%%%%%%%%%%%%
%%%%%%%%%%%%%%%%%%%%%%%%%%%%%%%%%%%%%%%%
\section{Formulation of the problem and relevant equations}

The relevant radiative transfer problem here implies calculating, at each spatial 
point of the (generally magnetized) astrophysical plasma model under consideration, the values of the diagonal and non-diagonal elements of the atomic density matrix corresponding to each atomic level of total angular momentum $J$, which quantify its overall population, the population imbalances between its magnetic sublevels, and the quantum coherences between each pair of them. The values of such density-matrix elements have to be consistent with the intensity, polarization and symmetry properties of the radiation field generated within the medium. This requires solving jointly the RT 
equations for the Stokes parameters and the statistical equilibrium equations for the density-matrix elements.
Once such a self-consistent solution is obtained it is straightforward to compute the emergent Stokes profiles via the formal solution of the Stokes-vector transfer equation for any desired line of sight (LOS). The aim of this section is to formulate such  equations with emphasis on the case of the IR triplet of Ca {\sc ii}.

\subsection{The multipolar components of the atomic density matrix}

The excitation state of the atomic system is described through the density matrix (Fano 1957). We consider the populations ($\rho_J(M, M)$), and coherences ($\rho_J(M, M')$) among the $(2J+1)$ magnetic sublevels $M$ of each given atomic level with total angular momentum $J$. Alternatively, we use the multipolar components of the atomic density matrix (Omont 1977):
\begin{equation}
{\rho^K_Q}(J)=\sum_{MM'}(-1)^{J-M}{\sqrt{2K+1}}
\left( \begin{array}{ccc}
J&J&K \\
M&-M^{'}&-Q
\end{array} \right)
{\rho_{J}}(M,M^{'}),
\end{equation}
where $K=0, 1, ..., 2J$, $Q=-K, -K+1, ..., K-1, K$, 
and the expression in brackets   
is the Wigner 3j-symbol (e.g., Brink \& Satchler 1968).
The $\rho^K_0$ elements with $K{\ne}0$ 
are real numbers measuring population imbalances between 
sublevels, while $\sqrt{2J+1}\rho^0_0$ is the total population of the level. 
The $\rho^K_{Q\ne 0}$ elements are combinations of the complex non-diagonal
components $\rho_J(M, M')$. Hermiticity of the density matrix ($\rho_J(M, M')=\rho_J(M', M)^*$)
implies $\rho^K_{-Q}=(-1)^Q[\rho^K_{Q}]^*$ ($*$ meaning complex conjugate).
In the absence of net circular polarization in the incident radiation  
$\rho_J(M, M)=\rho_J(-M, -M)$, i.e., atoms {\em align} but they do not {\em orientate}. 
Hence, odd-$K$ $\rho^K_{Q}$ elements vanish.
Therefore, the 5-level atomic model of Ca~{\sc ii} considered here (Fig.~1) requires $1+6+15+1+6=29$ 
$\rho^K_{Q}$ elements to fully characterize the excitation state of the atomic system.

\subsection{The transfer equation for the Stokes parameters}

The transfer of spectral line polarization is described by 
the following equations for the Stokes parameters 
\begin{subequations}\label{rte}
\begin{align}
\frac{\rm d}{{\rm d}s}I&=\epsilon_I-\eta_II-\eta_QQ-\eta_UU-\eta_VV, \label{rte1}
\displaybreak[0] \\
\frac{\rm d}{{\rm d}s}Q&=\epsilon_Q-\eta_QI-\eta_IQ-\rho_VU+\rho_UV, \label{rte2}
\displaybreak[0] \\
\frac{\rm d}{{\rm d}s}U&=\epsilon_U-\eta_UI+\rho_VQ-\eta_IU-\rho_QV, \label{rte3}
\displaybreak[0] \\
\frac{\rm d}{{\rm d}s}V&=\epsilon_V-\eta_VI-\rho_UQ+\rho_QU-\eta_IV.\label{rte4}
\end{align}
\end{subequations}
In the Hanle effect regime the magnetic splitting of $\sigma$ and $\pi$ components of 
the line transitions are negligible and the spectral line emissivity thus depends 
exclusively on the excitation state of the upper level of the transition (i.e., on the $\rho^K_Q(J_u)$ values).
In particular, linearly polarized emission is due to the presence of alignment
and coherences between atomic sublevels of the upper level of the transition. 
Moreover, since levels are not oriented, $\epsilon_V^{\rm line}\equiv 0$.
More explicitly (Landi Degl'Innocenti \& Landolfi 2004; 
see also Manso Sainz \& Trujillo Bueno 1999):
\begin{subequations}\label{epsilons}
\begin{multline}
  {\epsilon}_I^{\rm line}={\epsilon_{0}}\,{{\rho}^0_0}+
  {\epsilon_{0}}{w^{(2)}_{J_uJ_l}}
  \Big{\{}\frac{1}{2\sqrt{2}}(3\mu^2-1){{\rho}^2_0} -
  \sqrt{3}  \mu \sqrt{1-\mu^2} (\cos \chi
  {\rm Re}[{{{\rho}}^2_1}] - \sin 
  \chi {\rm Im}[{{{\rho}}^2_1}]) \\
+\frac{\sqrt{3}}{2} (1-\mu^2) (\cos
  2\chi \, {\rm Re}[{{{\rho}}^2_2}]-\sin 2\chi \, 
  {\rm Im}[{{{\rho}}^2_2}]) \Big{\}},
\end{multline}
\begin{multline}
 {\epsilon}_Q^{\rm line}=\,-\,{\epsilon_{0}}\,w^{(2)}_{J_uJ_l}\Big{\{}\frac{3}{2\sqrt{2}}(\mu^2-1) 
  {\rho}^2_0 -
  \sqrt{3}  \mu \sqrt{1-\mu^2} (\cos \chi
  {\rm Re}[{{\rho}}^2_1] - \sin 
  \chi {\rm Im}[{{\rho}}^2_1]) \\
- \frac{\sqrt{3}}{2} (1+\mu^2) (\cos
  2\chi \, {\rm Re}[{{\rho}}^2_2]-\sin 2\chi \, {\rm Im}[{{\rho}}^2_2]) \Big{\}},
\end{multline}
\begin{equation}
\epsilon_U^{\rm line}=\,-\,{\epsilon_{0}}\,w^{(2)}_{J_uJ_l}\sqrt{3} \,\Big{\{} \sqrt{1-\mu^2} ( \sin \chi
  {\rm Re}[{{\rho}}^2_1]+\cos \chi {\rm Im}[{{\rho}}^2_1]) 
 +\mu (\sin 2\chi \, {\rm Re}[{{\rho}}^2_2] + \cos 2\chi \, {\rm Im}[{{\rho}}^2_2]) \Big{\}},
\end{equation}
\end{subequations}
where $\epsilon_0=(h\nu/4\pi)A_{ul}{\phi_x}{\cal N}\sqrt{2J_u+1}$
(with $\cal N$ the total number of atoms per unit volume, $A_{ul}$ the Einstein coefficient for spontaneous emission and ${\phi_x}$ the Voigt profile, with $x=\nu_0 - \nu$ and $\nu_0$ the transition's frequency), 
$\theta=\arccos\ (\mu)$ and $\chi$ are the inclination and azimuth of the ray, and
$w^{(2)}_{J_uJ_l}$
is a numerical coefficient that depends only on $J_u$ and $J_l$ 
In particular,  $w^{(2)}_{1/2,3/2}=0$, $w^{(2)}_{3/2,3/2}=-2\sqrt{2}/5$, and $w^{(2)}_{3/2,5/2}=\sqrt{2}/10$
(see Table 10.1 in Landi Degl'Innocenti \& Landolfi 2004).

The absorption coefficient $\eta_I^{\rm line}$, 
and the dichroism coefficients $\eta_Q^{\rm line}$ and $\eta_U^{\rm line}$ are given by analogous expressions 
but with $\eta_0=(h\nu/4\pi)B_{lu}{\phi_x}{\cal N}\sqrt{2J_l+1}$ instead of $\epsilon_0$,
$w^{(2)}_{J_lJ_u}$ instead of $w^{(2)}_{J_uJ_l}$ ($w^{(2)}_{3/2,1/2}=\sqrt{2}/2$, $w^{(2)}_{5/2,3/2}=\sqrt{7}/5$), and with the $\rho^K_Q$
values of the {\it lower} level of the line transition instead of those of the upper level
(we neglect stimulated emissions). Likewise, since we are assuming that the atomic levels are not oriented, $\eta_V^{\rm line}\equiv 0$ (when neglecting the wavelength shifts between the $\sigma$ components caused by the Zeeman effect). Finally, the anomalous dispersion terms ($\rho^{line}_{Q,U}$) have the same functional dependence on the $\rho^K_Q$ values of the {\it lower} level, but with an antisymmetric dispersive dependence on frequency, i.e., $\rho^{line}_{Q,U}=\eta^{line}_{Q,U}\,{\psi_{\nu}}/{\phi_{\nu}}$, with ${\psi_{\nu}}$ the dispersion profile.

Additionally, we consider an unpolarized background continuum providing an extra absorption coefficient
$\eta_I^{\rm cont}$ and emissivity $\epsilon_{I}^{\rm cont}={\eta_{I}}^{\rm cont}B_{\nu}$ (where $B_{\nu}$ is the Planck function), a very good approximation toward the red part of the solar spectrum (e.g., Gandorfer 2000).
Therefore, in Eqs.~(3), $\eta_I=\eta_I^{\rm line}+\eta_I^{\rm cont}$, 
$\epsilon_I=\epsilon_I^{\rm line}+\epsilon_I^{\rm cont}$, while all other emission, dichroism,
and anomalous dispersion coefficients have only line contributions 
($\epsilon_Q=\epsilon_Q^{\rm line}$, $\epsilon_U=\epsilon_U^{\rm line}$, 
$\eta_Q=\eta_Q^{\rm line}$, $\eta_U=\eta_U^{\rm line}$, $\rho_Q=\rho_Q^{\rm line}$, 
$\rho_U=\rho_U^{\rm line}$). Note that $\epsilon_Q$ and $\eta_Q$ depend on both the population imbalances ($\rho^2_0$) and on the coherences ($\rho^2_Q$, with $Q=1,2$), while $\epsilon_U$ and $\eta_U$
depend {\it only} on the quantum coherences\footnote{In these expressions, the reference direction for Stokes $Q$ is the {\em perpendicular} to the plane formed by the propagation direction and the quantization Z-axis of total angular momentum}.

The $\rho^0_0$ elements
produce the dominant contribution to the Stokes $I$ profile, while 
the $\rho^2_Q$ elements (the {\em alignment} components)
contribute to the {\em linear} polarization signals, which
we quantify by the Stokes parameters $Q$ and $U$. 
%The $\rho^1_Q$ elements (the {\em orientation} components) contribute to the circular polarization, but they are negligible for the Ca {\sc ii} problem under consideration because there is no significant net circular polarization in the radiation that pumps the quiet Sun atoms. 
Although $\epsilon_V=\eta_V=0$ (because, as mentioned above, $\rho^1_Q=0$)  
according to Eq.~(3d) it is 
possible to generate Stokes $V$ in the spectral line due to the $-\rho_UQ+\rho_QU$
(second-order) terms. This contribution is however negligible compared to the longitudinal Zeeman effect
signal produced by the splitting of the $\sigma$-components of the emission profile
which we do not consider here. 

Since the degree of anisotropy of the radiation field in solar-like atmospheres is weak, the population imbalances and coherences of each $J$-level are only a small fraction of its overall population. Therefore, second-order terms can be safely neglected in the Stokes-vector transfer equation and
the relevant equations for calculating $Q/I$ and $U/I$ are
\begin{subequations}
\begin{align}
\frac{\rm d}{{\rm d}s}I & \approx \epsilon_I-\eta_I I, \\
\frac{\rm d}{{\rm d}s}Q & \approx [\epsilon_Q-\eta_QI] -\eta_I Q, \\
\frac{\rm d}{{\rm d}s}U & \approx [\epsilon_U-\eta_UI] -\eta_I U. 
\end{align}
\end{subequations}
Nevertheless, all the calculations of this paper were carried out by solving numerically the full Stokes vector
transfer equations (3) through the application of the formal solver mentioned in Appendix A. 
The complication with respect to the unmagnetized case considered by Manso Sainz \& Trujillo Bueno (2003{\natexlab{a}}) is that now the presence of a magnetic field of given strength $B$, inclination $\theta_B$ and azimuth $\chi_B$ at each grid point of the stellar atmosphere model under consideration implies that we have to consider also the quantum coherences ($\rho^2_Q$, with $Q{\ne}0$) between each pair of magnetic substates pertaining to each $J$ level, in addition to their individual populations ($\rho^2_0$). For this reason, not only Stokes $Q$ but also Stokes $U$ is now non-zero, in general.

The expressions above for the components of the emission vector
and propagation matrix (Eqs. 4 and paragraphs following them)
simplify considerably for several cases of practical interest. 
For a one-dimensional atmosphere, either unmagnetized, or permeated by a {\em microturbulent} and {\em isotropically distributed} magnetic field of strength $B$, or in the presence of a deterministic magnetic field with a fixed inclination but
with a random azimuth below the line's photon mean-free-path, or in the presence of a deterministic magnetic field with a fixed orientation but with a strength in the saturation regime of the Hanle effect, the quantum coherences (i.e. the real and imaginary parts of the $\rho^2_1$ and $\rho^2_2$ components) vanish
in a reference system whose quantization axis must be the parallel to the magnetic field vector 
for the deterministic field case and the the normal to the stellar surface for the three other cases:

\begin{equation}
  \epsilon^{\rm line}_I\,=\,\epsilon_0 \big{[}\rho^0_0+
	w_{J_uJ_\ell}^{(2)} \frac{1}{2\sqrt{2}} (3\mu^2-1)\rho^2_0\big{]},
\end{equation}
\begin{equation}
  \epsilon^{\rm line}_Q\,=\,\epsilon_0 w_{J_uJ_\ell}^{(2)} \frac{3}{2\sqrt{2}}(1-\mu^2)
	\rho^2_0,
\end{equation}
\begin{equation}
  \epsilon^{\rm line}_U\,=0,
\end{equation}
where the $\rho^K_0$ ($K=0,2$)
values are those of the {\it upper} level of the line transition under consideration,
and where the orientation of the ray is specified by the azimuthal angle $\chi$ and by 
$\mu={\rm cos}\theta$ (with $\theta$ the angle between the direction of the radiation beam 
and the quantization axis, which must be the magnetic field direction itself for the deterministic field case 
and the normal to the surface of the stellar 
atmospheric model for the two other cases). Likewise, the only non-zero elements of the propagation matrix are
$\eta_I$ and $\eta_Q$ which are given by
identical expressions (i.e. by $\eta_I=\epsilon_I$ and 
$\eta_Q=\epsilon_Q$),
but with $\eta_0$ instead of $\epsilon_0$,
$w^{(2)}_{J_lJ_u}$ instead of $w^{(2)}_{J_uJ_l}$ and with the $\rho^K_0$
values of the {\it lower} level of the line transition instead of those of the upper level
(for the case in which stimulated emissions are neglected).

Under such assumptions, and taking into account 
the low polarization level in solar-like atmospheres (i.e., that $\eta_Q/\eta_I\ll 1$ and 
$\epsilon_Q/\epsilon_I\ll 1$), 
an approximate formula can be derived to estimate the emergent
fractional linear polarization amplitude at the center of a strong spectral line 
(Trujillo Bueno 2003):
\begin{equation}
Q/I\,\approx\,{3\over{2\sqrt{2}}}(1-\mu^2)
[w_{J_uJ_\ell}^{(2)}\,\sigma^2_0({J_u})\,-\,w^{(2)}_{J_lJ_u}\,\sigma^2_0({J_l})],
\end{equation} 
where $\sigma^2_0=\rho^2_0/\rho^0_0$ must be evaluated at height in the model atmosphere where the
line-center optical distance along the line of sight is unity.
Its first term is due to {\em selective emission} of polarization components (caused by the population imbalances of the upper level), while its second term accounts for ``zero-field'' dichroism. Note that since $w^{(2)}_{1/2,3/2}=0$ and $w^{(2)}_{3/2,1/2}=\sqrt{2}/2$ the scattering polarization of ${\lambda}8662$ is exclusively due to ``zero-field'' dichroism.

\subsection{The statistical equilibrium equations}

General equations for the multivel atom model which neglects 
quantum interferences between the sublevels pertaining to different $J$-levels are derived in
Landi Degl'Innocenti \& Landolfi (2004). 
For the particular case of the 5-level atom of Fig.~1 the following 
rate equations for the density matrix elements follow:
\begin{align}
\begin{split}\label{see01}
\frac{\rm d}{{\rm d}t} \rho^0_0(1) =& -\biggl[\sum_{u=4}^5 B_{1u}J^0_0(1\rightarrow u) + \sum_{i\neq 1} C_{1i} \biggr] \rho^0_0(1)
 + A_{41}\rho^0_0(4) + \sqrt{2} A_{51}\rho^0_0(5) +
\sum_{i\neq 1} C_{i1} \sqrt{\frac{2J_i+1}{2}}  \rho^0_0(i),
\end{split}
\displaybreak[0] \\
\begin{split}\label{see02}
\frac{\rm d}{{\rm d}t} \rho^0_0(2) =& -\biggl[\sum_{u=4}^5 B_{2u}J^0_0(2\rightarrow u) + \sum_{i\neq 2} C_{2i} \biggr] \rho^0_0(2) 
- \sum_{Q'=-2}^2 \biggl( \frac{1}{\sqrt{2}} B_{24}J^2_{Q'}(2\rightarrow 4)
-\frac{2\sqrt{2}}{5} B_{25}J^2_{Q'}(2\rightarrow 5) \biggr) \rho^2_{Q'}(2) \\
& + \frac{1}{\sqrt{2}} A_{42}\rho^0_0(4) + A_{52}\rho^0_0(5) +
\sum_{i\neq 2} C_{i2} \frac{\sqrt{2J_i+1}}{2}  \rho^0_0(i),
\end{split}
\displaybreak[0] \\
\begin{split}\label{see03}
\frac{\rm d}{{\rm d}t} \rho^2_Q(2) =& -\biggl[ 2\pi{\rm i}\nu_{\rm L}g_2 Q + \sum_{u=4}^5 B_{2u}J^0_0(2\rightarrow u) + \sum_{i\neq 2} C_{2i} -D_2^{(2)}\biggr] \rho^2_Q(2) \\
& -  \biggl( \frac{1}{\sqrt{2}} B_{24}J^2_{-Q}(2\rightarrow 4)
-\frac{2\sqrt{2}}{5} B_{25}J^2_{-Q}(2\rightarrow 5) \biggr)(-1)^Q \rho^0_0(2) \\
& + \frac{1}{5} A_{52}\rho^2_Q(5) + \sum_{i=3, 5} C_{i2}^{(2)} \frac{\sqrt{2J_i+1}}{2}  \rho^2_Q(i),
\end{split}
\displaybreak[0] \\
\begin{split}\label{see04}
\frac{\rm d}{{\rm d}t} \rho^0_0(3) =& -\biggl[B_{35}J^0_0(3\rightarrow 5) + \sum_{i\neq 3} C_{3i} \biggr] \rho^0_0(3) 
- \sum_{Q=-2}^2 \frac{\sqrt{7}}{5} B_{35}J^2_Q(3\rightarrow 5) \rho^2_Q(3) \\
& + \sqrt{\frac{2}{3}} A_{53} \rho^2_Q(5) + \sum_{i\neq 3} C_{i3} \sqrt{\frac{2J_i+1}{6}}  \rho^0_0(i),
\end{split}
\displaybreak[0] \\
\begin{split}\label{see05}
\frac{\rm d}{{\rm d}t} \rho^2_Q(3) =& -\biggl[ 2\pi{\rm i}\nu_{\rm L}g_3 Q + B_{35}J^0_0(3\rightarrow 5) + \sum_{i\neq 3} C_{3i} -D_3^{(2)}\biggr] \rho^2_Q(3) 
- B_{35}(-1)^Q J^2_{-Q}(3\rightarrow 5)\frac{\sqrt{7}}{5} \rho^0_0(3) \\
& - \sum_{q,Q=-2}^2 B_{35}J^2_{q}(3\rightarrow 5) \biggl[
-\sqrt{\frac{5}{7}} \biggl( \begin{matrix} 2 & 2 & 2 \\ Q & -Q' & q \end{matrix} \biggr) (-1)^{Q'} \rho^2_{Q'}(3)
+ \frac{9}{2}\sqrt{\frac{3}{35}} \biggl( \begin{matrix} 2 & 4 & 2 \\ Q & -Q' & q \end{matrix} \biggr) (-1)^{Q'} \rho^4_{Q'}(3) \biggr] \\
& + \frac{2}{5}\sqrt{\frac{7}{3}} A_{53} \rho^2_Q(5) + \sum_{i=2, 5} C_{i3}^{(2)} \sqrt{\frac{2J_i+1}{6}}  \rho^2_0(i),
\end{split}
\displaybreak[0] \\
\begin{split}\label{see06}
\frac{\rm d}{{\rm d}t} \rho^4_Q(3) =& -\biggl[ 2\pi{\rm i}\nu_{\rm L}g_3 Q + B_{35}J^0_0(3\rightarrow 5) + \sum_{i\neq 3} C_{3i} -D_3^{(4)}\biggr] \rho^4_Q(3) \\
& - \sum_{q, Q'=-2}^2 B_{35} J^2_q(3\rightarrow 5) \biggl[ \frac{9}{2}\sqrt{\frac{3}{35}} 
\biggl( \begin{matrix} 4 & 2 & 2 \\ Q & -Q' & q \end{matrix} \biggr)(-1)^{Q'} \rho^2_{Q'}(3) 
+3\sqrt{\frac{11}{70}} \biggl( \begin{matrix} 4 & 4 & 2 \\ Q & -Q' & q \end{matrix} \biggr)(-1)^{Q'} \rho^4_{Q'}(3) \biggr].
\end{split}
\displaybreak[0] \\
\begin{split}\label{see07}
\frac{\rm d}{{\rm d}t} \rho^0_0(4) =& -\biggl[ \sum_{\ell=1}^2 A_{4\ell} + \sum_{i\neq 4} C_{4i} \biggr] \rho^0_0(4) 
+ \sum_{\ell=1}^2 B_{\ell 4} J^0_0(\ell \rightarrow 4) \sqrt{\frac{2J_\ell+1}{2}} \rho^0_0(\ell) \\
&  + \sum_{Q=-2}^2 B_{24}  J^2_Q(\ell \rightarrow 4) \rho^2_Q(2) + \sum_{i\neq 4} C_{i4} \sqrt{\frac{2J_i+1}{2}}  \rho^0_0(i),
\end{split}
\displaybreak[0] \\
\begin{split}\label{see08}
\frac{\rm d}{{\rm d}t} \rho^0_0(5) =& -\biggl[ \sum_{\ell=1}^3 A_{5\ell} + \sum_{i\neq 5} C_{5i} \biggr] \rho^0_0(5)
+ \sum_{\ell=1}^3 B_{\ell 5} J^0_0(\ell \rightarrow 5) \frac{\sqrt{2J_\ell+1}}{2} \rho^0_0(\ell) \\
&  + \sum_{Q=-2}^2 \biggl( -\frac{2\sqrt{2}}{5} B_{25} J^2_Q(2\rightarrow 5) \rho^2_Q(2)
   +\frac{\sqrt{42}}{10}  B_{35} J^2_Q(3\rightarrow 5) \rho^2_Q(3) \biggr) 
+  \sum_{i\neq 5} C_{i5} \frac{\sqrt{2J_i+1}}{2}  \rho^0_0(i),
\end{split}
\displaybreak[0] \\
\begin{split}\label{see09}
\frac{\rm d}{{\rm d}t} \rho^2_Q(5) =& -\biggl[ 2\pi{\rm i}\nu_{\rm L}g_5 Q + \sum_{\ell=1}^3 A_{5\ell} + \sum_{i\neq 5} C_{5i} -D_5^{(2)}\biggr] \rho^2_Q(5) 
 + \frac{1}{5}B_{25}J^0_0(2\rightarrow 5)\rho^2_Q(2) + \frac{\sqrt{21}}{5}B_{35}J^0_0(3\rightarrow 5)\rho^2_Q(3) \\
& +  -\frac{2\sqrt{2}}{5} B_{25} J^2_{-Q}(2\rightarrow 5) \rho^0_0(2)
                        +\sqrt{\frac{3}{100}} B_{35} J^2_{-Q}(3\rightarrow 5) \rho^0_0(3) \\
& + \sum_{Q'=-2}^2 \biggl( 2\sqrt{\frac{7}{5}} B_{25} J^2_{Q'}(2\rightarrow 5) \rho^2_{Q+Q'}(2) 
                        -\sqrt{\frac{3}{5}} B_{35} J^2_{Q'}(3\rightarrow 5) \rho^2_{Q+Q'}(3) \biggr)
                        \biggl( \begin{matrix} 2 & 2 & 2 \\ -Q & Q-Q' & -Q' \end{matrix} \biggr) \\
& + \sum_{i=2, 3} C_{i5}^{(2)} \frac{\sqrt{2J_i+1}}{2}  \rho^2_0(i).
\end{split}
\end{align}
In the previous equations $A_{u\ell}$ and $B_{\ell u}$ are the 
Einstein coefficients for spontaneous emission and absorption, respectively
(see Table~1).
$J^{K_r}_{Q_r}$ ($K_r=0, 2$; $Q_r=-K_r, ..., K_r$) are the radiation field
tensors introduced by Landi Degl'Innocenti (1984). They are integrals
over frequency and solid angle of the Stokes parameters (for their explicit
expressions appropiate for computational work see Manso Sainz \& Trujillo
Bueno 1999). In particular, $J^0_0$ is the mean intensity integrated over 
the absorption profile, while
\begin{equation}
{{{J}^2_0}}=\int {\rm d}{x} 
\oint \phi_x \frac{{\rm d} \vec{\Omega}}{4\pi}
\frac{1}{2\sqrt{2}} \left[(3\mu^2-1){{I_{x \vec{\Omega}}}}+
3(\mu^2-1){Q_{x \vec{\Omega}}}\right]\,.
\end{equation}
The contribution of other $J^K_Q$ components is very small here, 
where only the magnetic field breaks the symmetry of the radiation field.
Thus, $|J^2_{Q\neq 0}/J^0_0|\leq 10^{-4}$ throughout the whole outer
atmosphere in the five transitions relevant for our problem. 

$C_{\ell u}$ and $C_{u\ell}$ are the excitation and desexcitation collisional rates,
respectively (see Table~1), between levels $\ell$ and $u$,
while $C^{(2)}_{\ell u}$ and $C^{(2)}_{u\ell}$ are collisional transfer
rates for alignment between polarizable levels 2, 3, and 5. 
Derouich et al. (2007) have computed values for these transitions in Ca~{\sc ii}. 
Alternatively one can 
follow the approach suggested by Landi Degl'Innocenti \& Landolfi (2004),
and consider $C^{(2)}_{u\ell}=C_{\ell u}p$ where $p$ is just a numerical factor
depending on the quantum numbers of the transition. 
$D^{(K)}_{i}$ is the depolarization rate of the $K$-th multipole of level ${i}$
due to elastic collisions with neutral hydrogen. 
Theoretical values for Ca~{\sc ii} are given by Derouich et al. (2007),
who confirmed that such $D^{(K)}_{i}$ terms of the statistical equilibrium equations 
have only a very small impact on the atomic polarization of the 
metastable lower levels (cf., Manso Sainz \& Trujillo Bueno 2003{\natexlab{a}}). 

The first term on the r.h.s. of Eqs.~(\ref{see03}), (\ref{see05})-(\ref{see06}),
and (\ref{see09}) represents the relaxation of coherences in the presence of a 
magnetic field with Larmor frequency ${\nu}_{\rm L}=1.3996{\times}10^6B$ 
($B$ expressed in gauss and $\nu_{\rm L}$ in Hz), and $g_{J}$ the Land\'e factor of the $J$-level 
under consideration.
Equations~(\ref{see01})-(\ref{see09}) are expressed in a reference system
with the quantization axis along the magnetic field. 
The radiative and collisional terms remain formally invariant in any 
other reference system ---although clearly, their actual numerical values may change
due to the different values of $J^K_Q$ in the new reference system.
In an arbitrary reference system where the magnetic field has an inclination
$\theta_B$ with respect to the quantization axis and an azimuth $\chi_B$, 
the magnetic relaxation term reads:
\begin{multline}
2\pi{\rm i}\nu_{\rm L}g_JQ\rho^K_Q \rightarrow \\
2\pi{\rm i}\nu_{\rm L}g_J\biggl[ Q\cos\theta_B \rho^K_Q 
-\frac{1}{2}\sqrt{(K-Q)(K+Q+1)}\sin\theta_B{\rm e}^{{\rm i}\chi_B} \rho^K_{Q+1} 
+\frac{1}{2}\sqrt{(K+Q)(K-Q+1)}\sin\theta_B{\rm e}^{-{\rm i}\chi_B} \rho^K_{Q-1} \biggr], 
\end{multline}
where the term multiplying $\rho^K_{Q+1}$ ($\rho^K_{Q-1}$) is not considered when $Q=K$ ($Q=-K$).

The statistical equilibrium equations (\ref{see01})-(\ref{see09}) have
been derived under several hypotheses.
Collisions are considered under the impact approximation. 
The interaction time of the collision is negligible and hence, 
collisional and radiative terms can be added independently.
The colliders are assumed to be isotropically distributed 
(they are Maxwellian), and they cannot generate atomic polarization, 
though they can transfer it between levels\footnote{See Manso Sainz \& Trujillo Bueno (2009) for an 
interesting consequence of this fact for the case of permitted lines at EUV wavelengths 
(i.e., a polarization mechanism of EUV coronal lines and 
the possibility of mapping the magnetic fields of coronal loops).}.
Radiative rates have been derived within the framework
of the quantum theory of spectral line polarization described in Landi Degl'Innocenti and Landolfi (2004),
assuming that the radiation field is {\em spectrally flat} (i.e., 
lacking spectral structure), in spectral ranges of the order of the
energy separation between levels with coherences. 
This is an excellent approximation for the case under consideration since
we are considering relatively weak fields (less than $\sim$100~G) and
we neglect coherences between different $J$ levels.
Neglecting $J$-level interferences is a good approximation for 
modeling the Ca~{\sc ii} IR triplet and the core of the
H- and K-lines, but it cannot account for the general $Q/I$ pattern
of the UV doublet (see Stenflo 1980), which is beyond the scope of this work.
Finally, we neglect Doppler correlations due to the thermal movement
of the atoms. All these hypotheses together account for 
what is called complete frequency redistribution (CRD) in the laboraroty
frame, since correlations between the incident and scattered radiation 
fields are completly neglected.

We calculate the 29 $\rho^K_Q$ unknowns at each grid point of the chosen stellar atmosphere model assuming statistical equilibrium
\begin{equation}
{{\rm d}\over{\rm dt}}{\rho^K_Q(J)}\,=\,0.
\end{equation}
The resulting system of equations is linearly dependent and we substitute one of the equations, say, the one for the ground-level population, by the trace equation of the density matrix, which establishes the conservation of the number of particles
\begin{equation}
{\sum_{J_i}}\sqrt{2J_i+1}{\rho^0_0(J_i)}\,=\,1.
\end{equation}

We have developed a general multilevel radiative transfer computer 
program for the numerical solution of the non-LTE problem of the second kind, which we have applied in this investigation to obtain the self-consistent solution of the previous equations. See Appendix A for details.

\begin{table}
\caption{Data for 5-level atomic model of Ca {\sc ii}}
\centering
\begin{tabular}{cccccccc}
\hline
 $\lambda$ (\AA\ ) & $u$\footnote{Upper level level of the transition labeled in Fig.~1} & $\ell$\footnote{Lower level level of the transition in Fig.~1} & $A_{u\ell}$ (s$^{-1}$)\footnote{From NIST atomic spectra database {\tt http://www.nist.gov/physlab/data/asd.cfm}} & \multicolumn{4}{c}{$C_{u\ell}/N_e$ (s$^{-1}$cm$^{3}$)\footnote{From Shine \& Linsky (1974), with $N_e$ being the electron density.}} \\
 & & & & 3000~K & 5000~K & 7000~K & 9000~K \\
\hline
\multicolumn{8}{c}{Allowed transitions} \\
\hline
K    & 5 & 1 & $1.4\times 10^8$  & 3.60$\times 10^{-7}$ & 2.92$\times 10^{-7}$ & 2.59$\times 10^{-7}$ & 2.39$\times 10^{-7}$ \\
H    & 4 & 1 & $1.4\times 10^8$  & 3.60$\times 10^{-7}$ & 2.92$\times 10^{-7}$ & 2.59$\times 10^{-7}$ & 2.39$\times 10^{-7}$ \\
8498 & 5 & 2 & $1.11\times 10^6$ & 3.68$\times 10^{-7}$ & 2.93$\times 10^{-7}$ & 2.51$\times 10^{-7}$ & 2.23$\times 10^{-7}$ \\
8542 & 5 & 3 & $9.6\times 10^6$  & 1.59$\times 10^{-6}$ & 1.29$\times 10^{-6}$ & 1.12$\times 10^{-6}$ & 1.01$\times 10^{-6}$ \\
8662 & 4 & 2 & $1.06\times 10^7$ & 1.61$\times 10^{-6}$ & 1.31$\times 10^{-6}$ & 1.15$\times 10^{-6}$ & 1.04$\times 10^{-6}$ \\
\hline
\multicolumn{8}{c}{Forbiden transitions} \\
\hline
---  & 5 & 4 & ---               & 3.94$\times 10^{-7}$ & 3.05$\times 10^{-7}$ & 2.58$\times 10^{-7}$ & 2.27$\times 10^{-8}$ \\
---  & 4 & 3 & ---               & 2.69$\times 10^{-7}$ & 2.08$\times 10^{-7}$ & 1.76$\times 10^{-7}$ & 1.55$\times 10^{-7}$ \\
---  & 3 & 1 & ---               & 2.20$\times 10^{-7}$ & 1.71$\times 10^{-7}$ & 1.44$\times 10^{-7}$ & 1.27$\times 10^{-7}$ \\
---  & 3 & 2 & ---               & 5.51$\times 10^{-7}$ & 4.27$\times 10^{-7}$ & 3.61$\times 10^{-7}$ & 3.19$\times 10^{-7}$ \\
---  & 2 & 1 & ---               & 2.20$\times 10^{-7}$ & 1.71$\times 10^{-7}$ & 1.44$\times 10^{-7}$ & 1.27$\times 10^{-7}$ \\
\hline
\end{tabular}
\end{table}

%%%%%%%%%%%%%%%%%%%%%%%%%%%%%%%%%%%%%%%%
% RADIATIVE TRANSFER
%%%%%%%%%%%%%%%%%%%%%%%%%%%%%%%%%%%%%%%%
\section{The atmospheric and the atomic models}

High spatial and temporal resolution intensity images taken at the line center of strong absorption lines, like H$\alpha$ and Ca {\sc ii} 8542~\AA, show that the chromosphere of the quiet Sun is a highly inhomogeneous and dynamic medium (e.g., review by Rutten 2007). The spatial and dynamic complexity is such that it is obvious that the two one-dimensional (1D) atmospheric models we have chosen for this investigation (i.e., the ``hot" FAL-C model of Fontenla et al. 1993 and the ``cool" M-CO model of Avrett 1995) should only be considered as rough representations of the stratification of the kinetic temperature and density at two different phases of inter-network oscillations (see Fig. 2). Nevertheless, as we shall see below, our solution of the multilevel Hanle-effect problem for the IR triplet of Ca {\sc ii} in such semi-empirical models of the solar atmosphere allows us to reach the main goal of this paper, namely to demonstrate the diagnostic potential of the linear polarization signals produced by atomic polarization in the levels of the Ca {\sc ii} IR triplet. 

We consider a 5-level atomic model of Ca~{\sc ii} including the ground level,
the metastable ${}^2D_{3/2, 5/2}$ levels and the 
${}^2P_{1/2, 3/2}$ upper levels of the $H$ and $K$ lines (see Fig. 1). The excitation state
in such a model is given by 29 statistical tensor elements $\rho^K_Q$
($K=0$, 2, 4; $Q=-K, ..., K$), including the 
total populations of the five levels.
The radiative transition probabilities are those of Edl\'en \& Risberg (1956)
compiled at the NIST database\footnote{{\tt http://physics.nist.gov/PhysRefData/ASD}}. 
(Note that these transition probabilities for the infrared triplet are slightly larger than
those used by Manso Sainz \& Trujillo Bueno (2003a) for the unmagnetized case; correspondingly, 
the $Q/I$ values for the zero-field case are slightly larger here than in 
our previous work.)
Inelastic collisional rates and line broadening parameters 
have been taken from Shine \& Linsky (1974; see Table 1). Elastic depolarizing collisions 
are treated according to Lamb \& ter Haar (1971), which gives results similar to 
those obtained using the collisional rates given by Derouich et al. (2007). 
The collisional pumping of $K\neq 0$ statistical tensors are treated following
Landi Degl'Innocenti \& Landolfi (2004).

We have calculated the number density of Ca {\sc ii} ions at each atmospheric height of the model under consideration by solving the standard non-LTE radiative transfer problem for a realistic atomic model that includes also the bound-free transitions from all the bound levels of Ca {\sc ii} to the ground level of Ca {\sc iii}. With the resulting number densities of Ca {\sc ii} ions we then solved the non-LTE problem of the second kind for the atomic model of Fig. 1 (see the Appendix), which is sufficiently realistic for investigating the polarization signatures produced by atomic level polarization and the Hanle effect in the lines of the Ca {\sc ii} IR triplet. We point out that while $B_H$ (see Eq. 1) is of the order of a few gauss for the upper level of the 8498 \AA\ and 8542 \AA\ lines, it lies in the milligauss range for the lower (metastable) levels of the Ca {\sc ii} IR triplet.

\section{The unmagnetized reference case}

For symmetry reasons, 
in a one-dimensional model atmosphere without magnetic fields the only non-zero Stokes parameters are $I(\lambda)$ and $Q(\lambda)$ (choosing the polarization directions as in Eqs.~(\ref{epsilons})). 

Figure 3 shows the wavelength variation of the emergent fractional linear polarization ($Q/I$) in the Ca {\sc ii} IR triplet, calculated for a line of sight with $\mu=0.1$ in the ``hot" FAL-C model (middle panels) and in the ``cool" M-CO model (bottom panels). While $Q/I$ in the 8542 \AA\ and 8662 \AA\ lines are positive in both models, the $Q/I$ profile of the 8498 \AA\ line is positive in the ``hot" model but negative in the ``cool" model. Moreover, the 8498 \AA\ $Q/I$ profiles calculated in such atmospheric models have even different shapes, and their amplitudes are one order of magnitude smaller than those of the 8542 \AA\ and 8662 \AA\ lines. The (positive) $Q/I$ amplitudes of the 8542 \AA\ and 8662 \AA\ lines turn out to be significantly larger in the ``cool" model. This dependence of the amplitude of the $Q/I$ profiles on the chosen model atmosphere is not at all surprising, because the atomic polarization that anisotropic pumping processes induce in the atomic energy levels sensitively depends on the temperature structure of the model. In particular, the steeper the temperature gradient, the larger the anisotropy factor and hence, the induced atomic level polarization (e.g., see figure 4 in Trujillo Bueno 2001). The upper panels of Fig. 4 show the spatial variation of the ``degree of anisotropy" in each of the considered atmospheric models, indicating for each LOS characterized by $\mu$ the atmospheric height where the line-center optical depth is unity along the LOS. Note that at the height corresponding to a LOS with $\mu=0.1$ the anisotropy is larger in the M-CO model. 
The decrease of the anisotropy factor at very low optical depths
is due to the rapid widening of the $\phi_x$ profile with which
the radiation field is averaged (e.g., see Eq. (18) for the definition of the $J^2_0$
radiation field tensor component). This is a consequence of the 
temperature rising in the upper chromosphere-transition region. 
The two lower panels of Fig. 4 show the behavior of the fractional atomic alignment of the levels of Fig. 1 having $J>1/2$ (i.e., the two metastable levels, which are the lower levels of the Ca {\sc ii} IR triplet, and the upper level of the K-line). The fact that the sign of the $Q/I$ profile of $\lambda8498$ 
(i.e., the weakest line of the Ca {\sc ii} IR triplet) is very sensitive to the thermal structure of the lower chromosphere can be easily understood by combining the information provided in the lower panels of Fig. 4 and the approximate Eddington-Barbier formula (8).

Fig. 3 suggests that the scattering polarization of the Ca {\sc ii} IR triplet can be used as a sensitive thermometer of the ``quiet" regions of the solar chromosphere, as already noted by Manso Sainz \& Trujillo Bueno (2001). In this respect, a particularly useful quantity is the center-limb variation of the $Q/I$ line-center amplitudes (see Fig. 5). As seen in this figure, the most noteworthy feature is that in the ``hot" FAL-C model the (positive) $Q/I$ amplitude of the 8498 \AA\ line is larger at $\mu{\approx}0.2$ than at $\mu{\approx}0.1$ and that it becomes negative for line of sights with $\mu$ values significantly smaller than 0.1.

\section {The Hanle effect in the Ca {\sc ii} IR triplet}

The aim of this section is to show how the Hanle effect modifies the emergent linear polarization of the IR triplet of Ca {\sc ii}. To this end, we consider the following two geometries. 

\subsection {Forward scattering geometry}

In the absence of magnetic fields and horizontal atmospheric inhomogeneities the polarization of the atomic levels do not produce any linear polarization in the spectral line radiation observed at disc center ($\mu=1$; see Fig.~5). The same is true in the presence of a vertical magnetic field, because in a one-dimensional stellar atmospheric model the vertical direction coincides with the symmetry axis of the anisotropic radiation field that induces atomic level polarization. However, in the presence of an inclined magnetic field the symmetry of the problem is broken and, as a result, the ensuing atomic level polarization can generate linear polarization even for a line of sight with $\mu=1$. In this case, the linear polarization is created by the Hanle effect of the inclined field (see a demonstrative observational example in Trujillo Bueno et al. 2002).

Figure 6 shows the fractional linear polarization of the IR triplet of Ca {\sc ii} generated in forward scattering geometry by the Hanle effect of a horizontal magnetic field in the FAL-C (top panels) and M-CO (bottom panels) models. In the (``hot") FAL-C model, the linear polarization amplitudes are at the level of $10^{-4}$ when the magnetic strength of the horizontal field is similar or larger than only 0.1 gauss, with $Q/I>0$ (i.e., parallel to the horizontal magnetic field) for the 8498 \AA\ line and with $Q/I<0$ (i.e., perpendicular to the horizontal magnetic field) for the 8542 \AA\ and 8662 \AA\ lines. In the (``cool") M-CO model $Q/I>0$ in the 8542 \AA\ and 8662 \AA\ lines, with amplitudes again at the $10^{-4}$ level for $B\,{\gtrsim}\,0.1$ gauss, while the shape of $Q/I$ in the 8498 \AA\ line is very peculiar (i.e., similar to the typical observational signature of the transverse Zeeman effect) and with its maximum amplitude one order of magnitude smaller. Figure 7 shows the results of our RT calculations for the case of a magnetic field inclined by only $30^{\circ}$. Note that now, $U/I{\ne}0$. At present, detection of these weak linear polarization signals is challenging, though feasible. But they will become straigtforward with future large-aperture solar telescopes such as the Advanced Technology Solar Telescope (Keil, Rimmele \& Wagner 2009) or the European Solar Telescope (Collados 2008).

\subsection {Close to the limb geometry}

In forward scattering geometry, symmetry imposes that linear polarization vanishes if the magnetic field is inclined but has a random azimuth distribution, unlike in the fixed inclination and azimuth case just considered. However, for observations away from the solar disk center ($\mu{<}1$), scattering may produce linear polarization in the emergent spectral line radiation even for magnetic fields with a uniformly distributed azimuth within the spatio-temporal resolution element of the observations. This is illustrated in Fig. 8, which shows the linear polarization amplitudes of the Ca {\sc ii} IR triplet in the FAL-C model (top panels) and in the M-CO model (bottom panels) vary with the magnetic strength ($B$) and the inclination ($\theta_B$) of the assumed random-azimuth field for a close-to-the-limb observation ($\mu=0.1$). 

Figure 8 shows how sensitive the $Q/I$ line-center amplitude of the 8498 \AA\ line is to the model's thermal structure, and its very peculiar behavior in the sub-gauss range. The 8542 \AA\ and 8662 \AA\ lines are mostly sensitive to magnetic fields between 0.001 and 0.1 gauss, although the 8542 \AA\ line is also weakly sensitive to fields between 1 and 10 gauss. The 8542 \AA\ line enters into the saturation regime of the upper-level Hanle effect for fields stronger than $\sim$10 gauss, while just 0.1 gauss saturate the 8662 \AA\ line. On the contrary, the 8498 \AA\ line is highly sensitive in the whole regime 0.001--10 gauss. As seen in Fig. 3, in the absence of magnetic fields the linear polarization amplitude of the 8498 \AA\ line for a LOS with $\mu=0.1$ is one order of magnitude smaller than those corresponding to the 8542 \AA\ and 8662 \AA\ lines. However, the three Ca {\sc ii} lines may show similar $Q/I$ amplitudes if there is a magnetic field of the order of only 0.1 gauss in the atmospheric region where the linear polarization of the 8498 \AA\ line originates.

Figure 9 shows the emergent $Q/I$ amplitudes averaging the $Q/I$ profiles of both atmospheric models with a weight of 0.55 for the FAL-C profiles and 0.45 for the M-CO profiles. These weights are similar to those used by Avrett (1995) to simultaneously reproduce the intensity profiles observed in the 4.6 $\mu{\rm m}$ lines of CO and in the H \& K lines of Ca {\sc ii}, and to those used by Holzreuter et al. (2006) to fit the $Q/I$ profile of the Ca {\sc ii} K-line observed in a quiet regions very close to the solar limb. These type of fits obtained by mixing a cool and a hot component should not be considered as a realistic representation of the chromospheric conditions, but as further indication of the complexity of the (time-dependent) three-dimensional thermal structure of the solar chromosphere. Nevertheless, this two-component model shows that the $Q/I$ observations of the IR triplet of Ca {\sc ii} reported by Stenflo et al. (2000) are compatible with very weak magnetic fields in the ``region of formation" of the 8498 \AA\ line (i.e., in the lower chromosphere) and with fields stronger than 10 gauss (and inclined by about $30^{\circ}$) in the ``regions of formation" of the 8542 \AA\ and 8662 \AA\ lines (i.e., in the upper chromosphere).

Stokes $U$ is zero if the magnetic field has a random-azimuth distribution, and it is non-zero if the atmosphere is permeated by a magnetic field vector with a well-defined inclination and azimuth. This is illustrated in the upper panels of Fig. 10, which corresponds to the FAL-C model. In the absence of magnetic fields, only $Q/I$ is non-zero (see the dashed lines), but $U/I$ becomes significant even at fields strengths as low as 0.005 gauss. Note that for this case (B=0.005 gauss), characterized by a horizontal magnetic field pointing towards the observer (triangles), perpendicular to the LOS (squares) and away from the LOS (diamonds), the Stokes $Q/I$ profiles of the Ca {\sc ii} IR triplet lines are always positive, while $U/I$ changes its sign. The dotted lines of Fig. 10 show the case of a 100~gauss horizontal field pointing towards the observer (i.e., a strength for which the three IR lines of Ca {\sc ii} are in the saturation regime of the upper-level Hanle effect). The bottom panels of Fig. 10 show the results of similar calculations for the cool M-CO model. In this model the $Q/I$ profile of the 8498 \AA\ line changes its sign and shape when the magnetic field azimuth changes from $\chi_B=0^{\circ}$ (or $\chi_B=180^{\circ}$) to $\chi_B=90^{\circ}$   

\subsection {Hanle-effect diagrams}

A Hanle-effect diagram  shows the line-center amplitudes of $Q/I$ and $U/I$ varying with the strength and the azimuth of the magnetic field 
corresponding to a given line-of-sight and inclination of the magnetic field vector. Figure 11 shows  restricted Hanle-effect diagrams corresponding to the configurations considered in  Fig. 10 (close-to-the-limb $\mu=0.1$ observation). In particular, the thin solid lines show how linear polarization varies with the magnetic strength, for a horizontal magnetic field pointing towards the observer. We recall that 50 gauss are needed to completely depolarize the 8498 \AA\ line, while just 10 gauss are sufficient for the 8542 \AA\ line and only 0.1 gauss for the 8662 \AA\ line (see Fig. 8). Thick solid lines show, for the case of a 0.005 gauss horizontal field, how the $Q/I$ and $U/I$ line center amplitudes vary as the orientation of the magnetic field vector changes from pointing towads the observer (triangles), to being perpendicular to the line of sight (squares) to pointing away from the observer (diamonds). Comparing the upper and lower panels we see the differences due to the different thermal structures of the FAL-C and M-CO models.

Figures 12 and 13 show full Hanle-effect diagrams obtained from the emergent Stokes 
$Q/I$ and $U/I$ profiles calculated for a close-to-the-limb ($\mu=0.1$) line of sight
in the ``hot" and ``cool" atmospheric models, respectively. Each panel corresponds to a fixed
inclination of the magnetic field vector. Each of the four thick lines in any given panel correspond to a fixed strength of the magnetic field with azimuth ranging from $\chi_B=0^{\circ}$ to $\chi_B=180^{\circ}$ (solid line), and from $\chi_B=180^{\circ}$ to $\chi_B=360^{\circ}$ (dashed line). Thin lines correspond to magnetic fields ranging between 0 G and 1000 G for a fixed $\chi_B$. They show the pattern of depolarization ($Q/I$ decreases) and rotation of the direction of linear polarization ($U/I$ appears) characteristic of the Hanle effect in a close to the limb scattering geometry. 

Finally, Figs. 14 and 15 show Hanle-effect diagrams for a disc center observation ($\mu=1$). Note that in this forward scattering geometry the Hanle effect of a magnetic field inclined with respect to the solar local vertical creates linear polarization.

%%%%%%%%%%%%%%%%%%%%%%%%%%%%%%%%%%%%%%%%
%%%%%%%%%%%%%%%%%%%%%%%%%%%%%%%%%%%%%%%%
% CONCLUSIONS
%%%%%%%%%%%%%%%%%%%%%%%%%%%%%%%%%%%%%%%%
%%%%%%%%%%%%%%%%%%%%%%%%%%%%%%%%%%%%%%%%
\section{Conclusions}

We have investigated theoretically the sensitivity of the scattering polarization $Q/I$ and $U/I$ profiles of the Ca {\sc ii} IR triplet to the presence of magnetic fields through the Hanle effect. To this end, we have applied a multilevel radiative transfer 
code (see Appendix A) for treating the transfer of spectral line polarization due to scattering in weakly magnetized stellar atmospheres. We find that the Ca {\sc ii} 8542 \AA\ and 8662 \AA\ lines are mostly sensitive to fields between 0.001 and 0.1 G, the former being also weakly sensitive to fields between 1 and 10 G. Otherwise the linear polarization signals are dependent only on the magnetic field geometry, but not to its intensity. The 8498 \AA\ line is highly sensitive in the whole regime 0.001--10 G, but its scattering polarization amplitude is significantly lower in the absence of magnetic fields. 

Probably, the most interesting line of the Ca {\sc ii} IR triplet because of its diagnostic potential is the strongest one: 
${\lambda}8542$. Its linear polarization is sensitive to the orientation of the magnetic field in the chromosphere of the quiet Sun, though not too much to its strength unless $B<1$ G there. Although its scattering polarization amplitude depends on the thermal structure of the solar chromosphere, the emergent $Q/I$ and $U/I$ profiles do not change sign between the (``hot") FAL-C model and the (``cool") M-CO model, unlike ${\lambda}8498$, which is very sensitive to the physical conditions of the lower chromosphere. The thermal sensitivity of the $\lambda8662$ line (which is sensitive only to milligauss field strengths) is similar to that of the 8542 \AA\ line. Therefore, spatial fluctuations in the sign of the emergent $U/I$ profiles of the Ca {\sc ii} 8542 \AA\ and 8662 \AA\ lines may be ascribed more safely to spatial variations in the azimuth of the chromospheric magnetic field than is the case for ${\lambda}8498$.

More  generally, the emergent $Q/I$ and $U/I$ profiles are produced by the joint action of atomic level polarization, and the Hanle and transverse Zeeman effects. In this paper we have considered only the action of atomic level polarization and the Hanle effect. These effects dominate the emergent linear polarization profiles for inclined magnetic fields with strengths weaker than $B_0$, where the $B_0$ value depends on the scattering geometry. 

In the forward scattering geometry of a disk center observation ($\mu=1$) the linear polarization of the Ca {\sc ii} IR triplet is dominated by the Hanle effect if the magnetic field is weaker than about 10 G. In fact, while the contribution of the transverse Zeeman effect is negligible for $0{\le}B{<}10$ G the Hanle effect creates fractional linear polarization signals of the order of $10^{-4}$ already for horizontal magnetic fields as weak as 0.1 G (see Figs. 6 and 7). For magnetic strengths $10{\lesssim}B{\lesssim}100$ G the contribution of the transverse Zeeman effect to the linear polarization observed at the solar disk center should not be neglected. Detection of $Q/I$ and $U/I$ disk center signals caused either by the Hanle effect alone (if $B{<}10$ G) or by the joint action of the Hanle and transverse Zeeman effects (if $10{\lesssim}B{\lesssim}100$ G) would require very high polarimetric sensitivity together with a spatial and temporal resolution sufficient to at least resolve partially the magnetic field azimuth. 

Given the weakness of the theoretical scattering polarization signals (${\lesssim}10^{-4}$ at $\mu=1$; ${\lesssim}10^{-3}$ at $\mu=0.1$), the most favourable geometry for observing them is in quiet regions close to the limb. In this geometry the emergent linear polarization of the Ca {\sc ii} IR triplet is non-zero even if the magnetic field has a random azimuth within the spatio-temporal resolution element of the observation (see Figs. 8 and 9). Moreover, for magnetic strengths sensibly weaker than 100 G the contribution of the transverse Zeeman effect to the emergent linear polarization is expected to be smaller than that caused by atomic level polarization. For stronger fields (e.g., $100{\lesssim}B{\lesssim}500$ G) the wings of the emergent $Q/I$ and $U/I$ profiles may show sizable features produced by the transverse Zeeman effect, but the line-center amplitudes would still be dominated by atomic level polarization. 

Finally, we point out that information on the relative amplitudes and signs of the scattering polarization profiles of the Ca {\sc ii} IR triplet would be very useful to help constrain the thermal and magnetic structure of the quiet solar chromosphere. In principle, to determine the structure of the magnetic field via Stokes inversion techniques is possible, although it would be easier if one is able to infer first the atmospheric thermal structure from the observed intensity spectrum.  
The $Q/I$ and $U/I$ signals we have studied here 
should also be exploited to investigate the reliability of three-dimensional models of the solar chromosphere resulting from holistic magnetohydrodynamic (MHD) simulations of the solar atmosphere. To this end, it suffices with confronting spectropolarimetric observations of the Ca {\sc ii} IR triplet with synthetic Stokes profiles obtained by solving the non-LTE problem of the 2nd kind in 3D snaphot models resulting from such MHD simulations.

\appendix 

{\bf The Numerical Solution of the non-LTE Problem of the Second Kind}

Modeling the 
spectral line polarization produced 
by atomic level polarization and its
modification by the Hanle effect requires calculating,
for multilevel systems, 
the excitation and ionization state
of chemical species of given abundance that is
consistent with both the intensity
and polarization of the radiation field generated
within the (generally magnetized) plasma
under consideration.
This so-called non-LTE problem of the 2nd kind (cf., Landi Degl'Innocenti \& Landolfi 2004) 
is a very involved non-local and non-linear radiative transfer problem 
which requires solving the statistical equilibrium equations for the elements
of the atomic density matrix and the Stokes-vector transfer equation
for each of the allowed transitions in the multilevel model.
Once such a self-consistent excitation state is known throughout the medium, it is then
straightforward to solve the transfer equation for any desired line of sight 
in order to obtain the emergent Stokes profiles to be compared with
spectropolarimetric observations. To this end, we developed a general 
computer program, which we will describe in greater detail 
in a future publication along with the efficient and accurate 
radiative transfer methods on which it is based (for a first advance see the workshop 
contribution by Manso Sainz \& Trujillo Bueno 2003{\natexlab{b}}; see also Manso Sainz 2002). 
In this Appendix we present only a 
brief summary of the code with emphasis on its application to solve the problem of scattering polarization 
and the Hanle effect in the IR triplet of ionized calcium.

A summary of the numerical approach is as follows:
\begin{enumerate}
\item The equations are formulated within the spherical tensors
representation of the density matrix and radiation field tensor.
There are $(2J+1)(J+1)$ unknowns $\rho^K_Q$ with $K$ even, for each
level with integer angular momentum $J$, and $(2J+1)J$ unknowns for each
level with semi-integer angular momentum $J$.
\item The radiative transfer equations for the Stokes parameters 
are integrated along short characteristics using 
the quasi-parabolic DELO method (DELOPAR; Trujillo Bueno 2003).
\item The iterative corrections for the unknowns (the statistical tensors
$\rho^K_Q$) are calculated applying a suitable generalization to the multilevel atom case 
of the fast iterative methods described in Trujillo Bueno \& Manso Sainz (1999).
This implies writing down the statistical equilibrium equations
taking explicitly into account the contribution of the diagonal
components of the $\Lambda$ operator and linearizing according 
to Eqs.~(53) and (54) of Trujillo Bueno (2003).
\item The iterative scheme can speeded up further 
through polynomial Ng-acceleration (e.g., Auer 1987).
\end{enumerate}

The number density of the ion under consideration is computed in non-LTE.
At each point in the atmosphere we solve the statistical equilibrium equations 
for the multipolar components of the atomic density matrix  
plus the conservation of particles equation to calculate the excitation state
of the atomic system. We calculate the radiation field at each point in the model atmosphere
for each of the allowed radiative transitions in the model atom 
by formal integration of the radiative transfer equations for the Stokes
parameters applying the DELOPAR method.
Then, radiation field tensors $J^K_Q$ are calculated and used to work out
the new excitation state of the atomic system and so on, iteratively. Figure 16 
shows the convergence behavior of the above-mentioned iterative method for the
numerical solution of the (Ca {\sc ii}) non-LTE problem of the second kind.

\acknowledgments Financial support by the Spanish Ministry of Science and Innovation 
through project AYA2007-63881 (Solar Magnetism and High-Precision 
Spectropolarimetry) is gratefully acknowledged.

\newpage

%
%FIGURES
%

\begin{figure}
\plotone{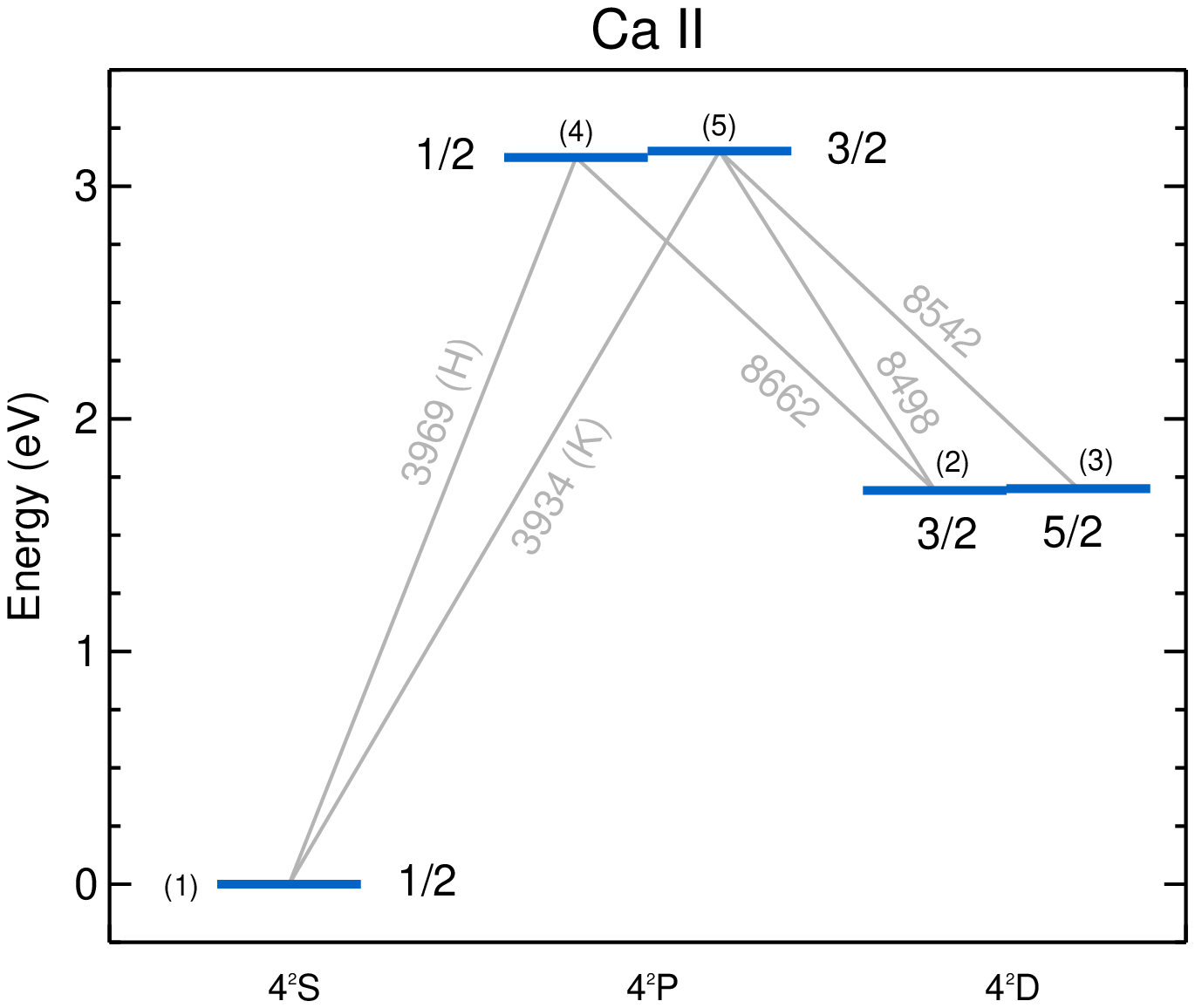}
\caption{The atomic model for Ca {\sc ii}. Labels indicate the total angular momentum of the level and its numbering (between parenthesis) in Eqs.~(\ref{see01})-(\ref{see09}). Solid lines connecting levels show allowed transitions and their labes their wavelength in \AA.
\label{fig1}}
\end{figure}

\begin{figure}
\plotone{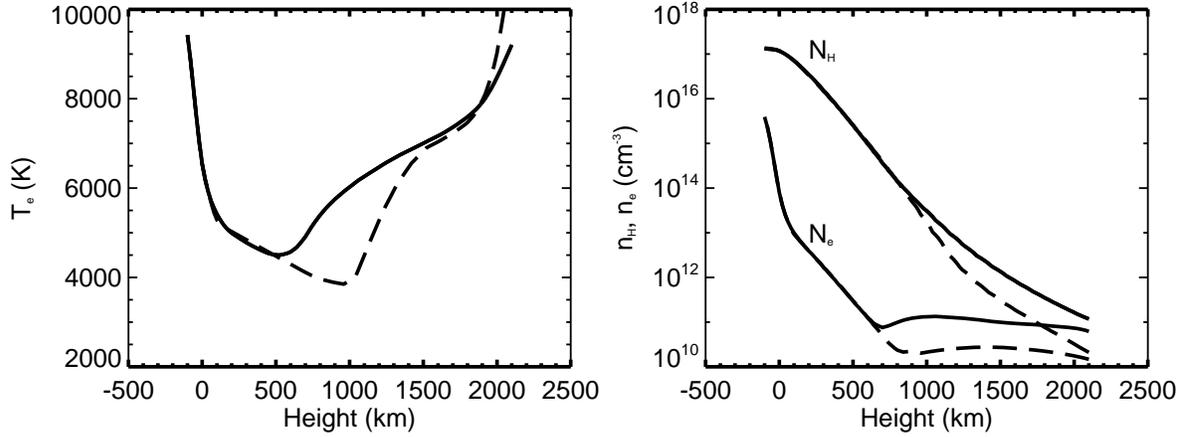}
\caption{The atmospheric models. Variation with height of the kinetic temperature (left panel), hydrogen and electron density (right panel) in the FAL-C model (solid lines), and M-CO model (dashed lines). Note in the left panel that the ``lower chromosphere'' of the M-CO model is relatively cool while it is relatively hot in FAL-C.
\label{fig2}}
\end{figure}

\begin{figure}
\plotone{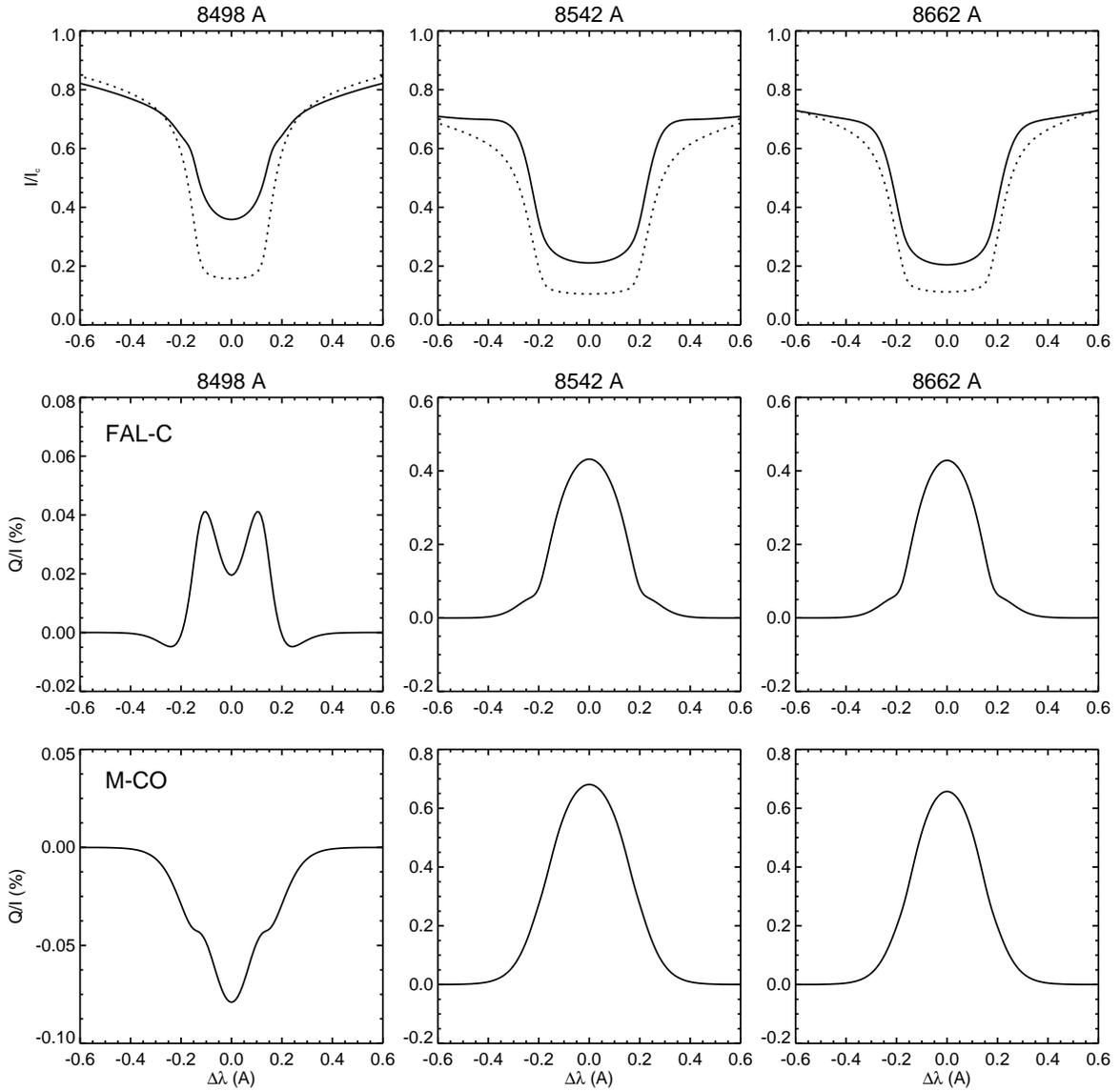}
\caption{The unmagnetized case. 
Emergent $Q/I$ profiles of the Ca {\sc ii} IR triplet calculated for a line of sight with $\mu=0.1$ in the ``hot" FAL-C model (middle panels) and in the ``cool" M-CO model (bottom panels). The positive $Q$-direction is parallel to the nearest limb. Note that the sign of the $Q/I$ profile of the 8498 \AA\ line is positive in the ``hot" model and negative in the ``cool" model.
The uppermost panel shows the emergent intensity profiles computed in the FAL-C model (solid lines) and in the M-CO model (dotted lines). 
\label{fig3}}
\end{figure}

\begin{figure}
\plotone{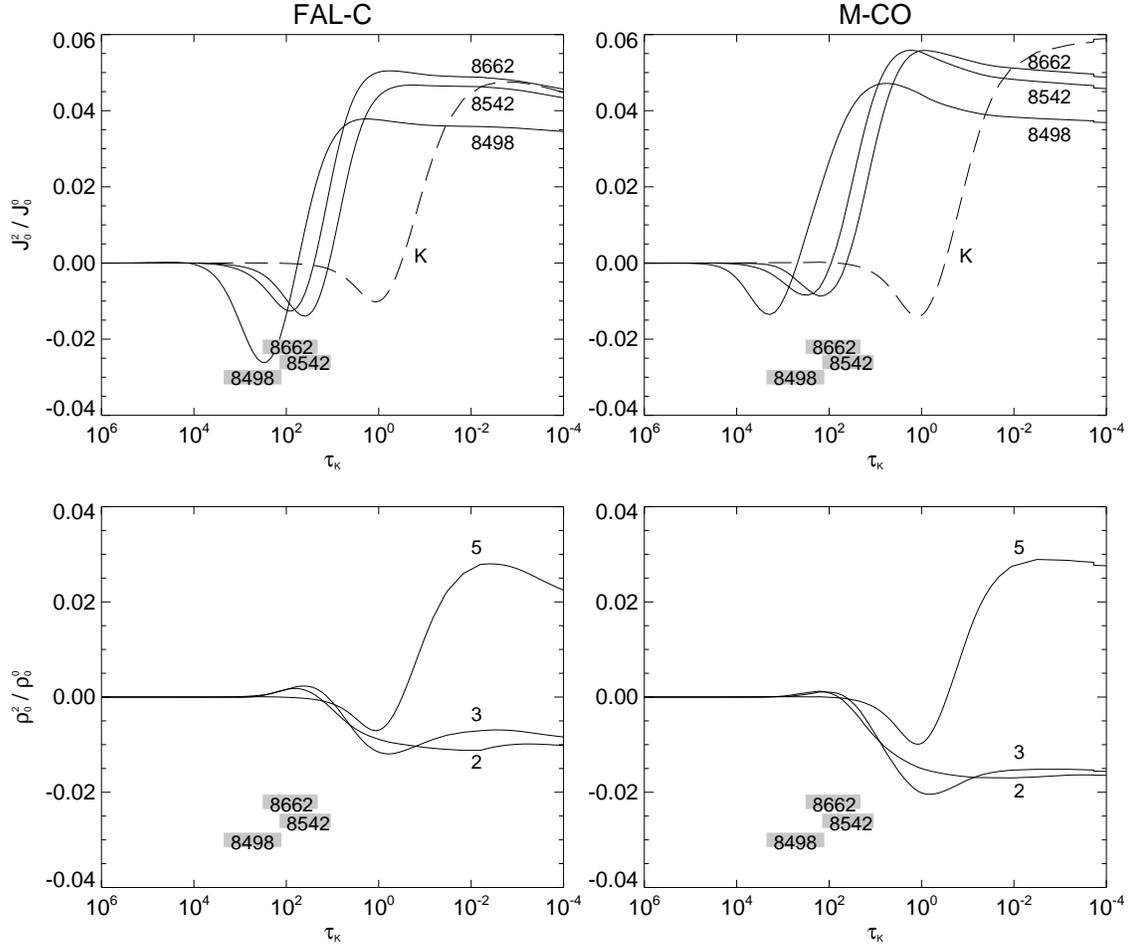}
\caption{The variation with the (vertical) K-line optical thickness of the degree of anisotropy in each of the lines of the Ca {\sc ii} model of Fig. 1 after obtaining the self-consistent solution in the FAL-C model (left panel) and in the M-CO model (right panel). For each of the IR triplet lines the corresponding thick horizontal segment indicate where its line-center optical depth is unity for line of sights going from $\mu=1$ (the lowest $\tau_{\rm K}$ of each segment) to $\mu=0.1$ (the largest $\tau_{\rm K}$ of each segment). 
\label{fig4}}
\end{figure}

\begin{figure*}
\plotone{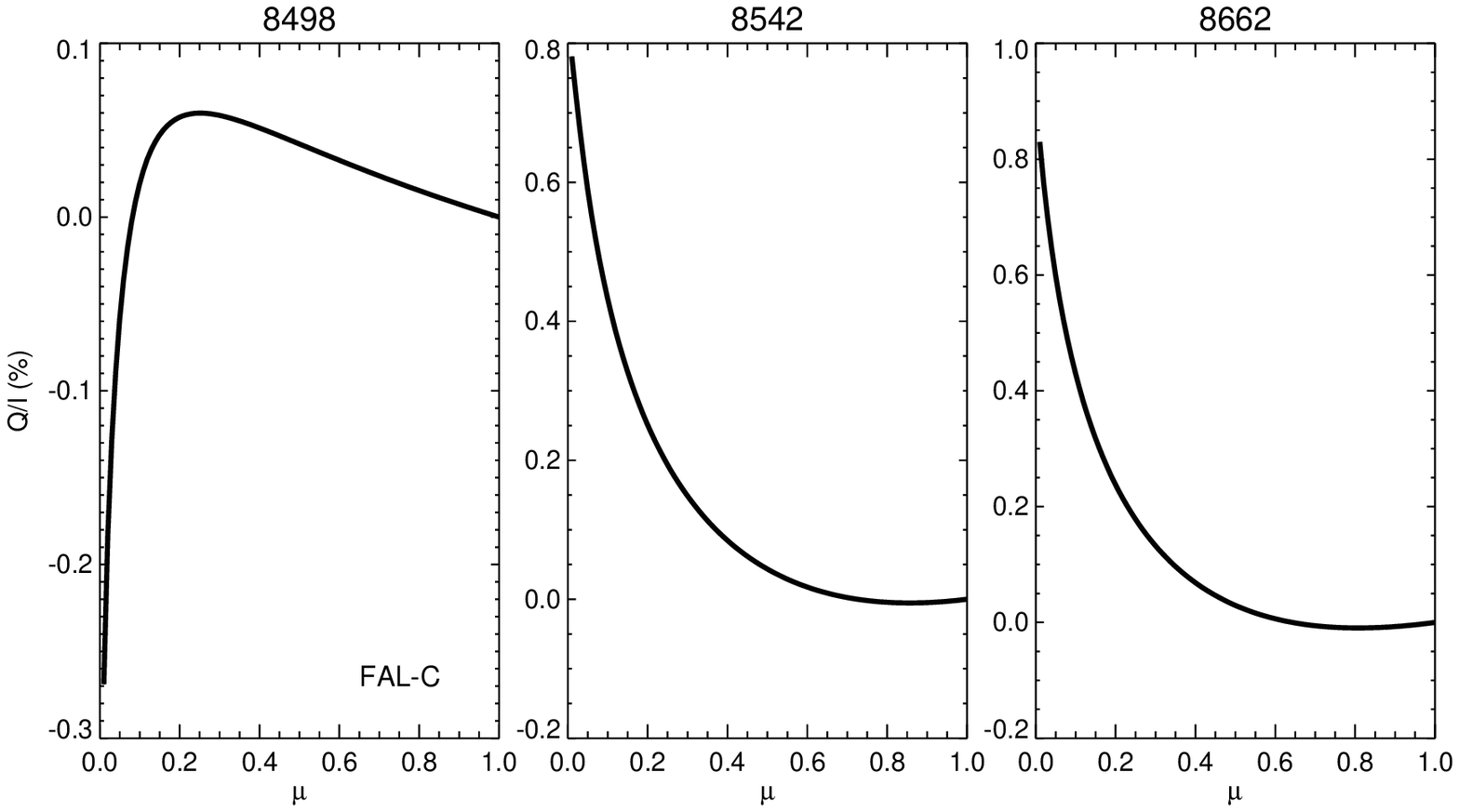}
\plotone{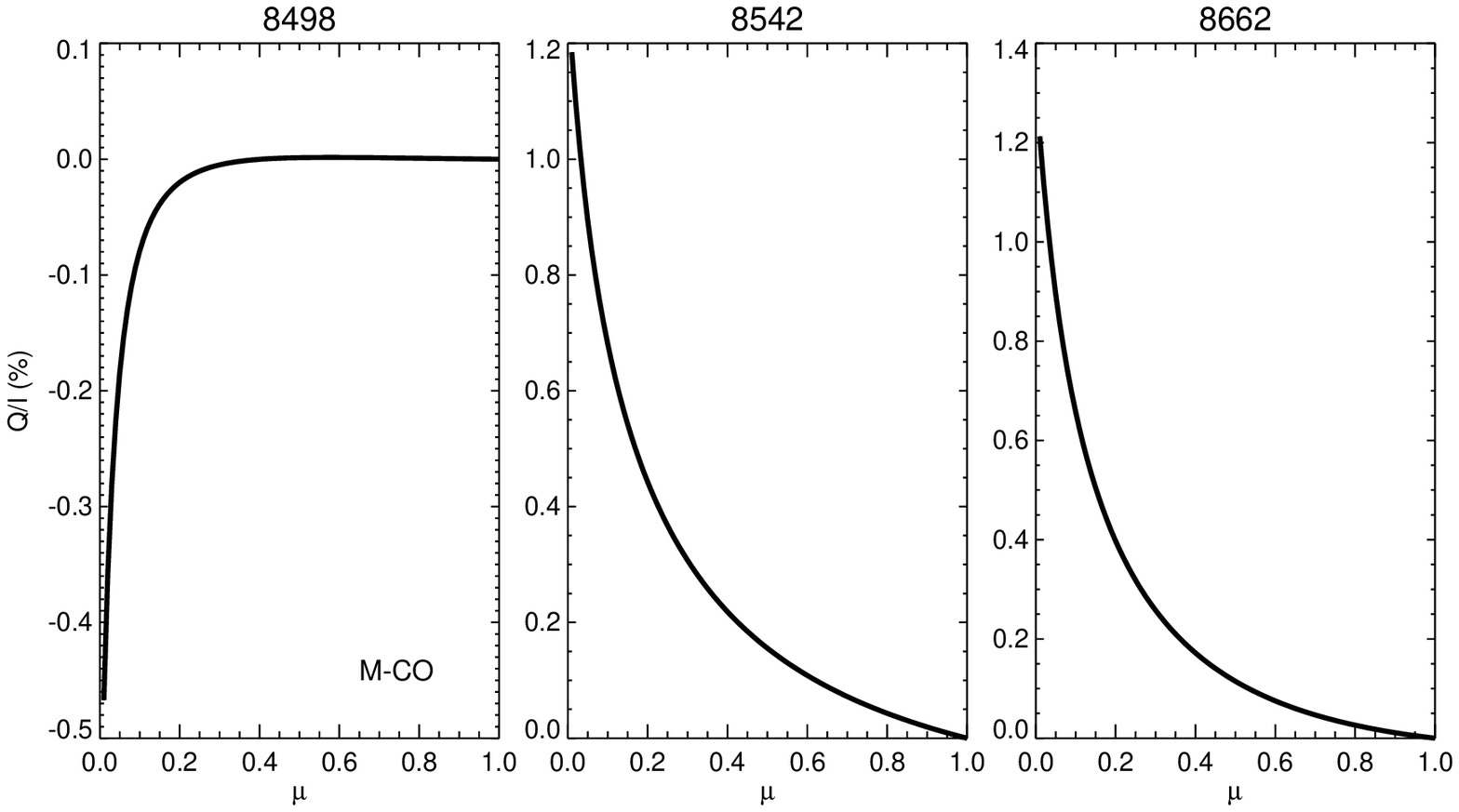}
\caption{The unmagnetized case. 
Center-to-limb variation of the line-center scattering polarization amplitudes in the Ca {\sc ii} IR triplet
produced by atomic level polarization in the ``hot" FAL-C model (top panels) and in the ``cool" M-CO model (bottom panels). The positive $Q$-direction is parallel to the nearest limb. Note that for relatively small $\mu$ values the $Q/I$ of $\lambda8498$ is negative for both model atmospheres. 
\label{fig5}}
\end{figure*}

\begin{figure*}
\plotone{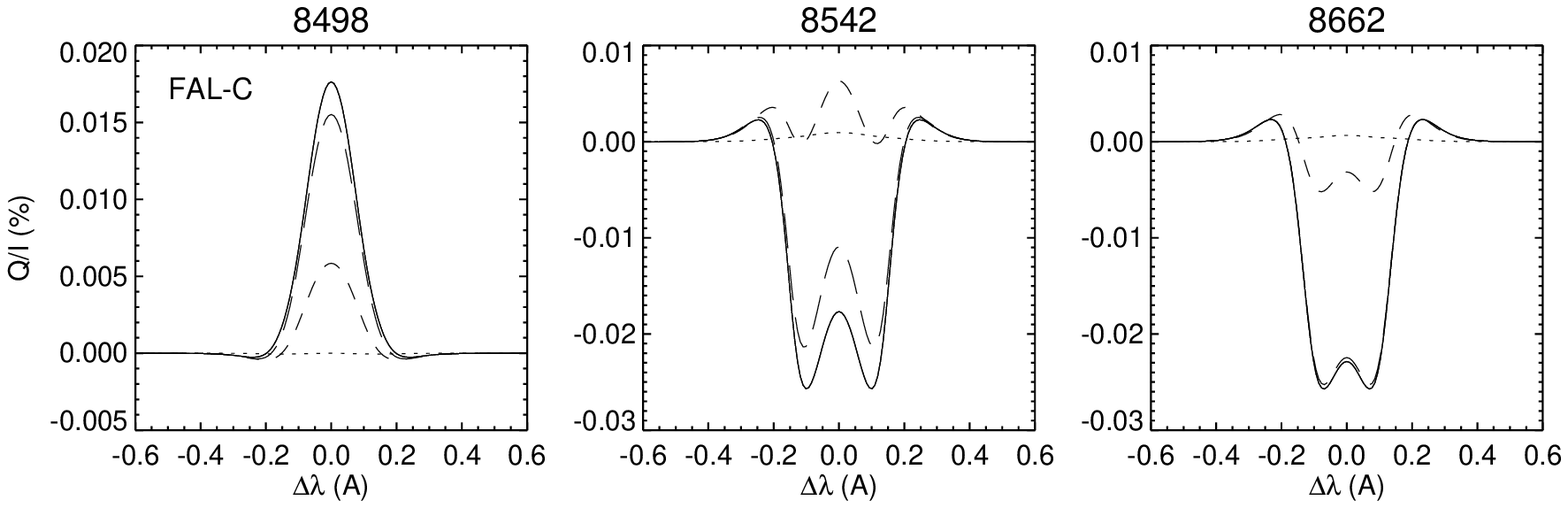}
\plotone{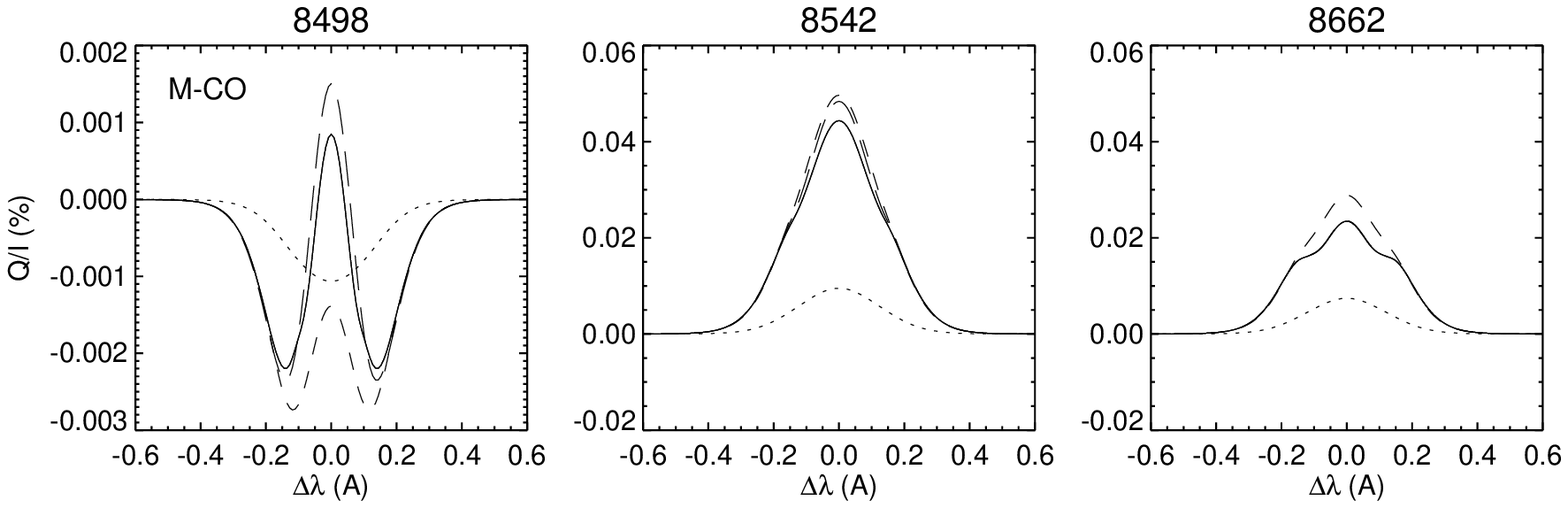}
\caption{The Hanle effect of a horizontal magnetic field in forward scattering. 
The emergent $Q/I$ profiles of the Ca {\sc ii} IR triplet calculated for a line of sight with $\mu=1$ (i.e., disk center observation) assuming the presence of a horizontal magnetic field of 100 G (solid lines), 0.1 G (long-dashed line), 0.01 G (short-dashed line), and 0.001 G (dotted line)
in the ``hot" FAL-C model (top panels) and in the ``cool" M-CO model (bottom panels). The positive $Q$-direction is along the magnetic field. 
\label{fig6}}
\end{figure*}

\begin{figure*}
\plotone{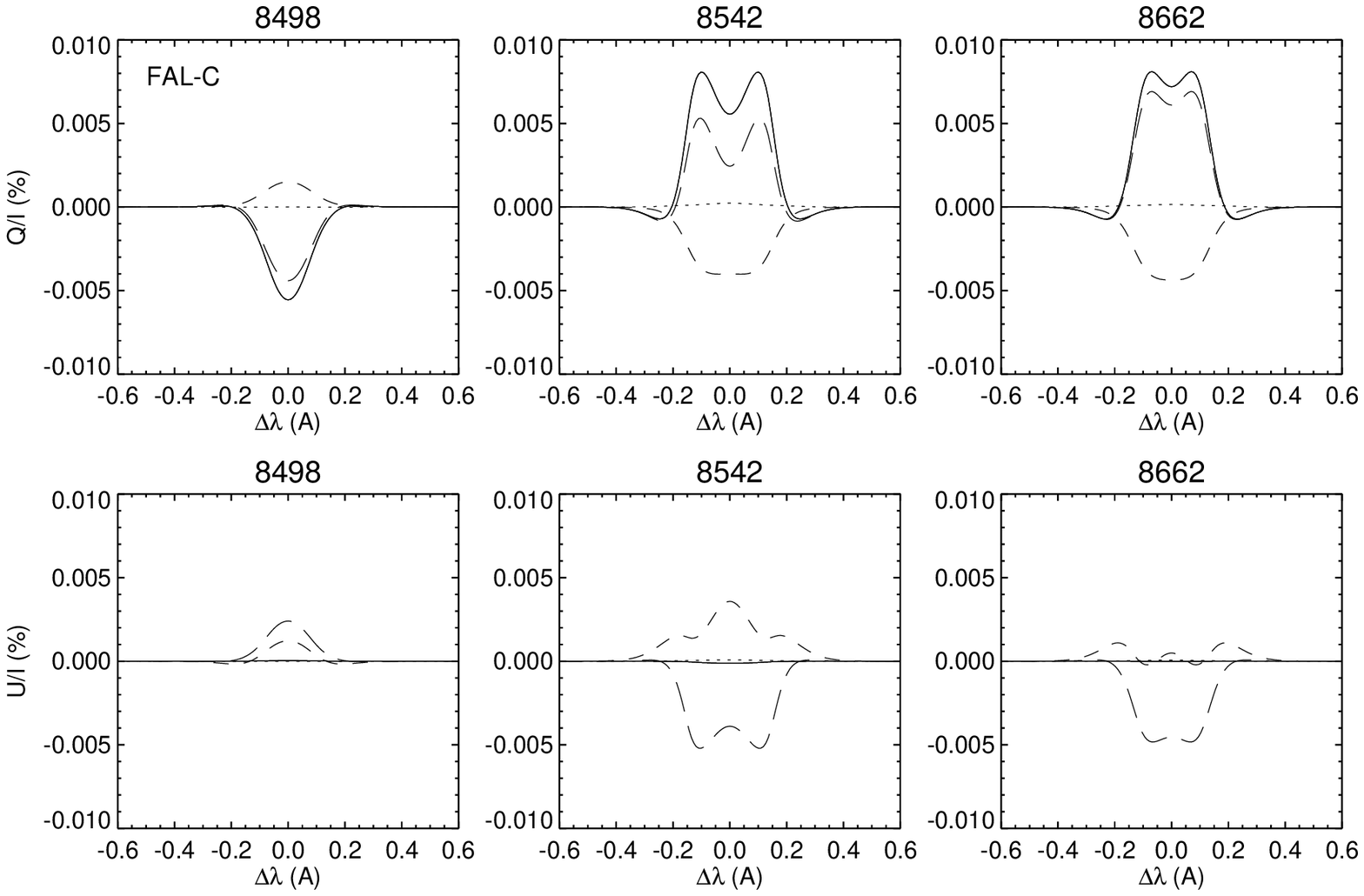}
\plotone{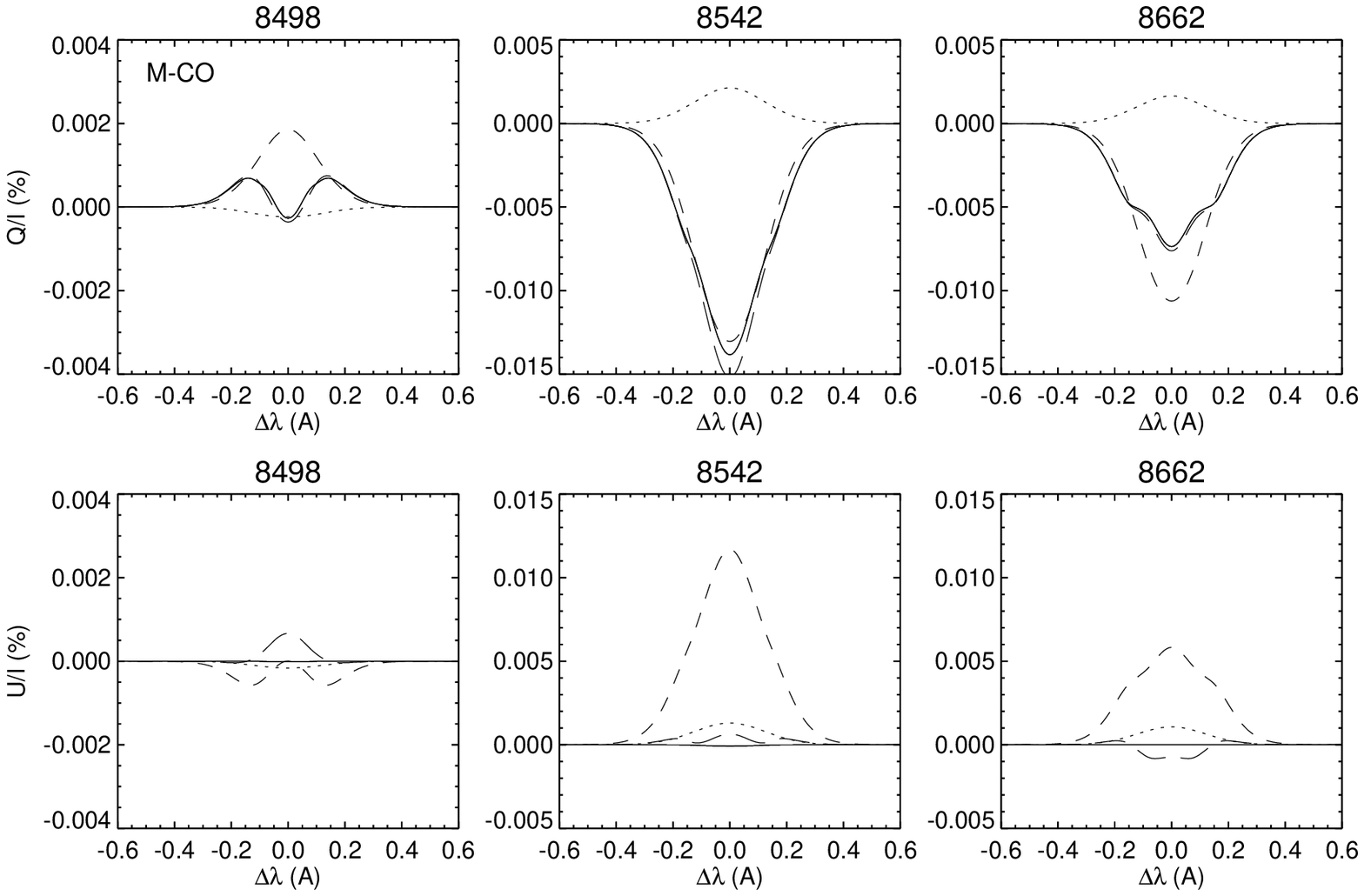}
\caption{The Hanle effect of a non-horizontal magnetic field in forward scattering. 
As in Fig. 6, but for a magnetic field inclined by $30^{\circ}$. Note that in this case Stokes $U$ is not zero. 
\label{fig7}}
\end{figure*}

\begin{figure*}
\plotone{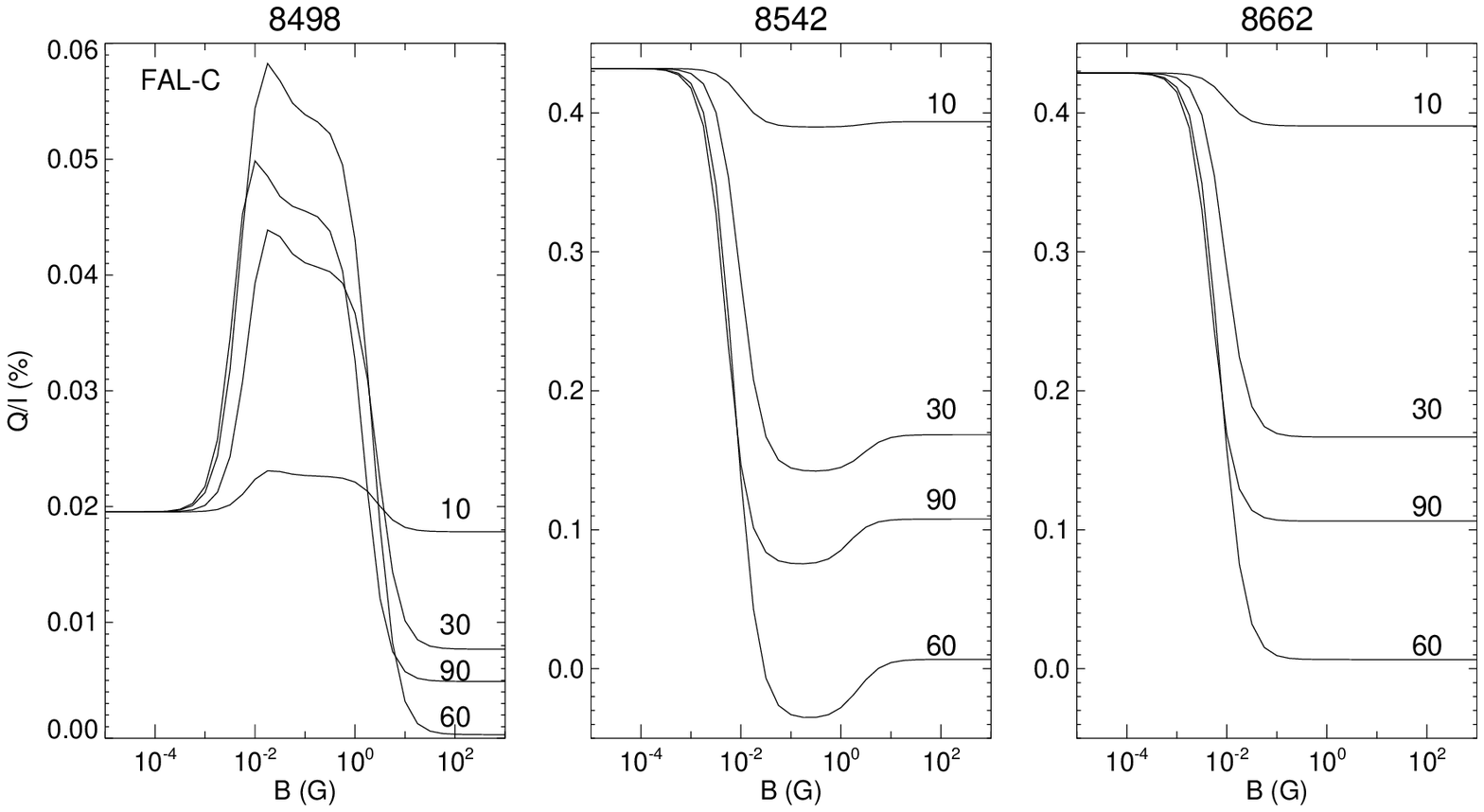}
\plotone{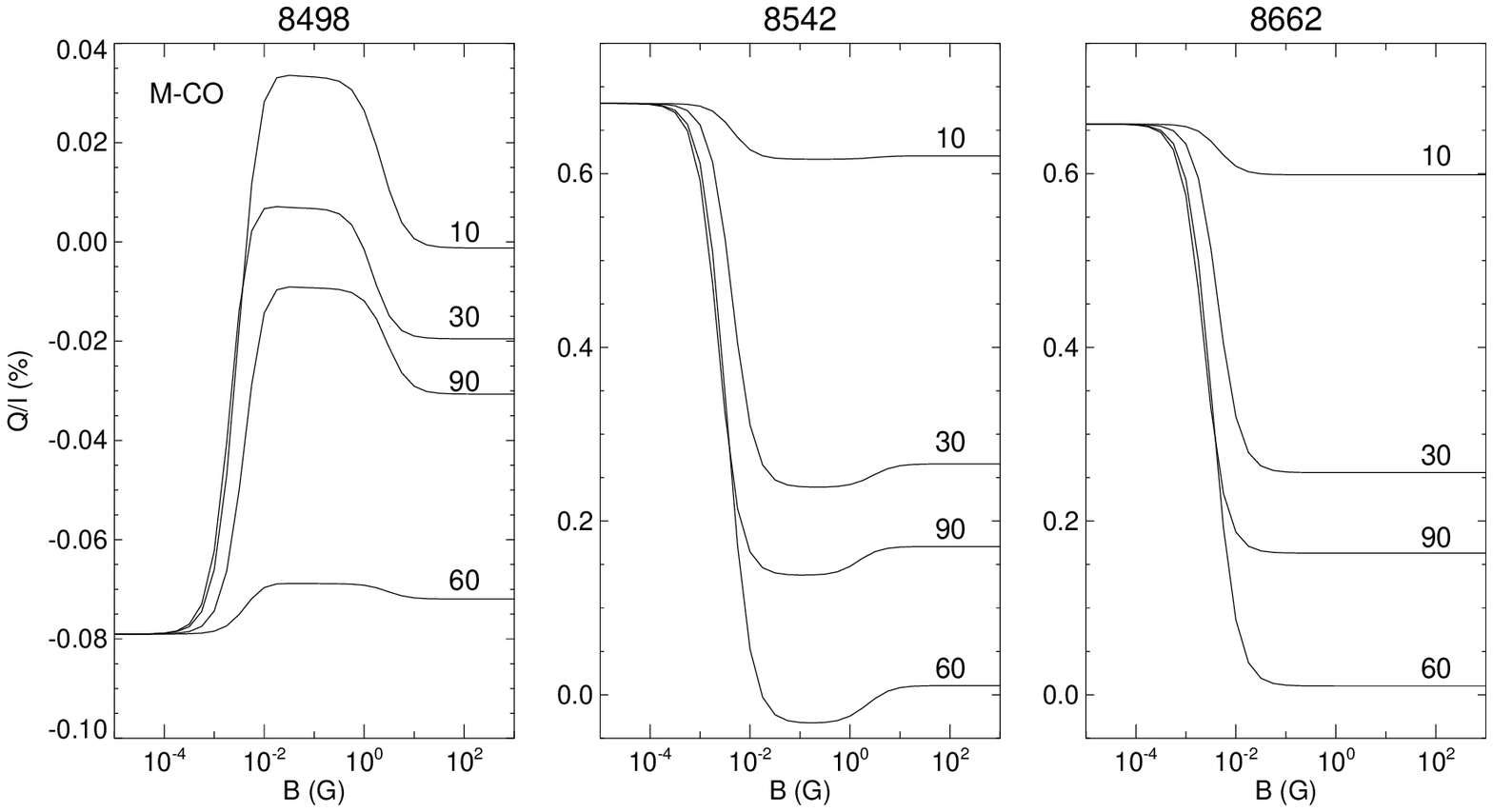}
\caption{The Hanle effect for a close to the limb line of sight.
The emergent $Q/I$ line-center amplitudes of the Ca {\sc ii} IR triplet calculated for a line of sight with $\mu=0.1$ in the ``hot" FAL-C model (top panels) and in the ``cool" M-CO model (bottom panels), assuming a magnetic field with the indicated inclination and a uniformly distributed azimuth within the spatio-temporal resolution element.  
The positive $Q$-direction is the parallel to the nearest limb. 
\label{fig8}}
\end{figure*}

\begin{figure}
\plotone{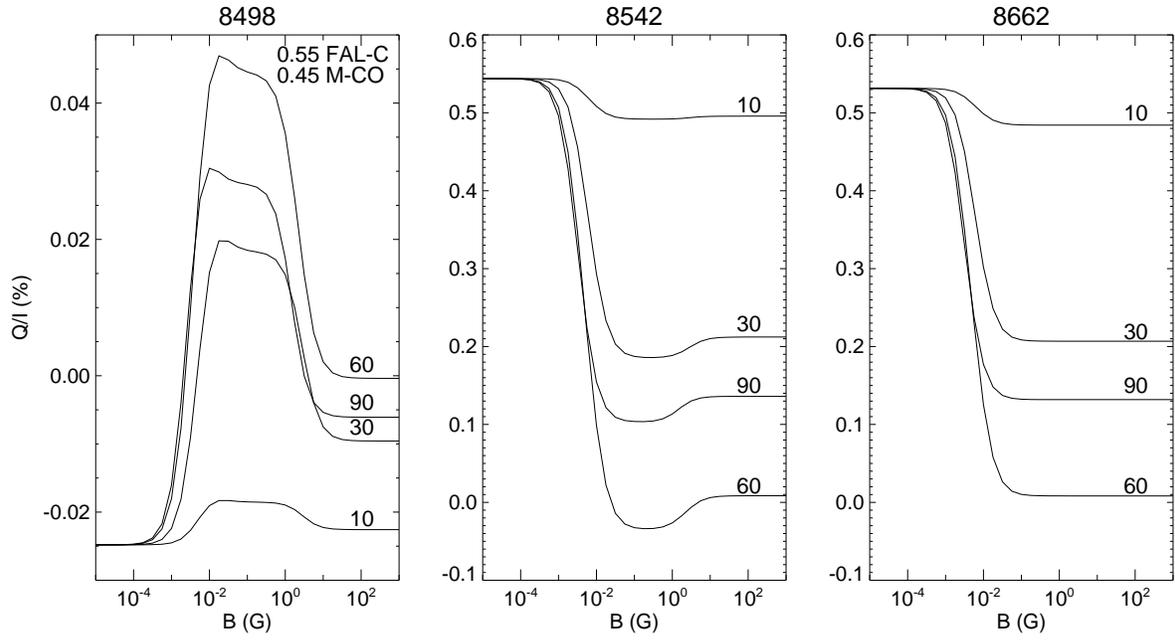}
\caption{The Hanle effect for a close to the limb line of sight.
The emergent $Q/I$ line-center amplitudes of the Ca {\sc ii} IR triplet calculated for a line of sight with $\mu=0.1$ assuming that 55\% of the radiation originates 
in the ``hot" FAL-C model and 45\% in the ``cool" M-CO model, assuming a magnetic field with the indicated inclination and a uniformly distributed azimuth. The positive $Q$-direction is the parallel to the nearest limb. 
\label{fig9}}
\end{figure}

\begin{figure*}
\plotone{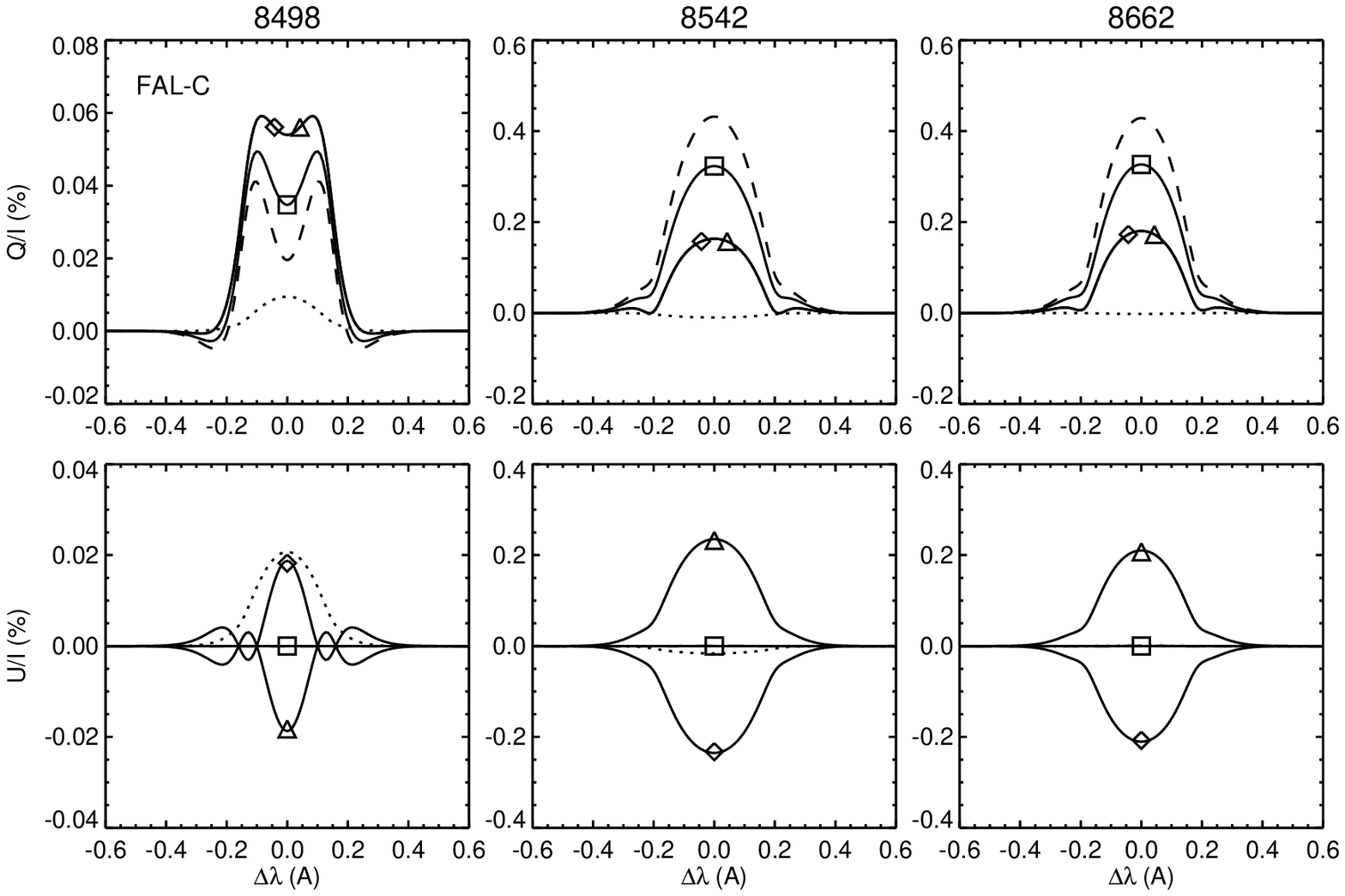}
\plotone{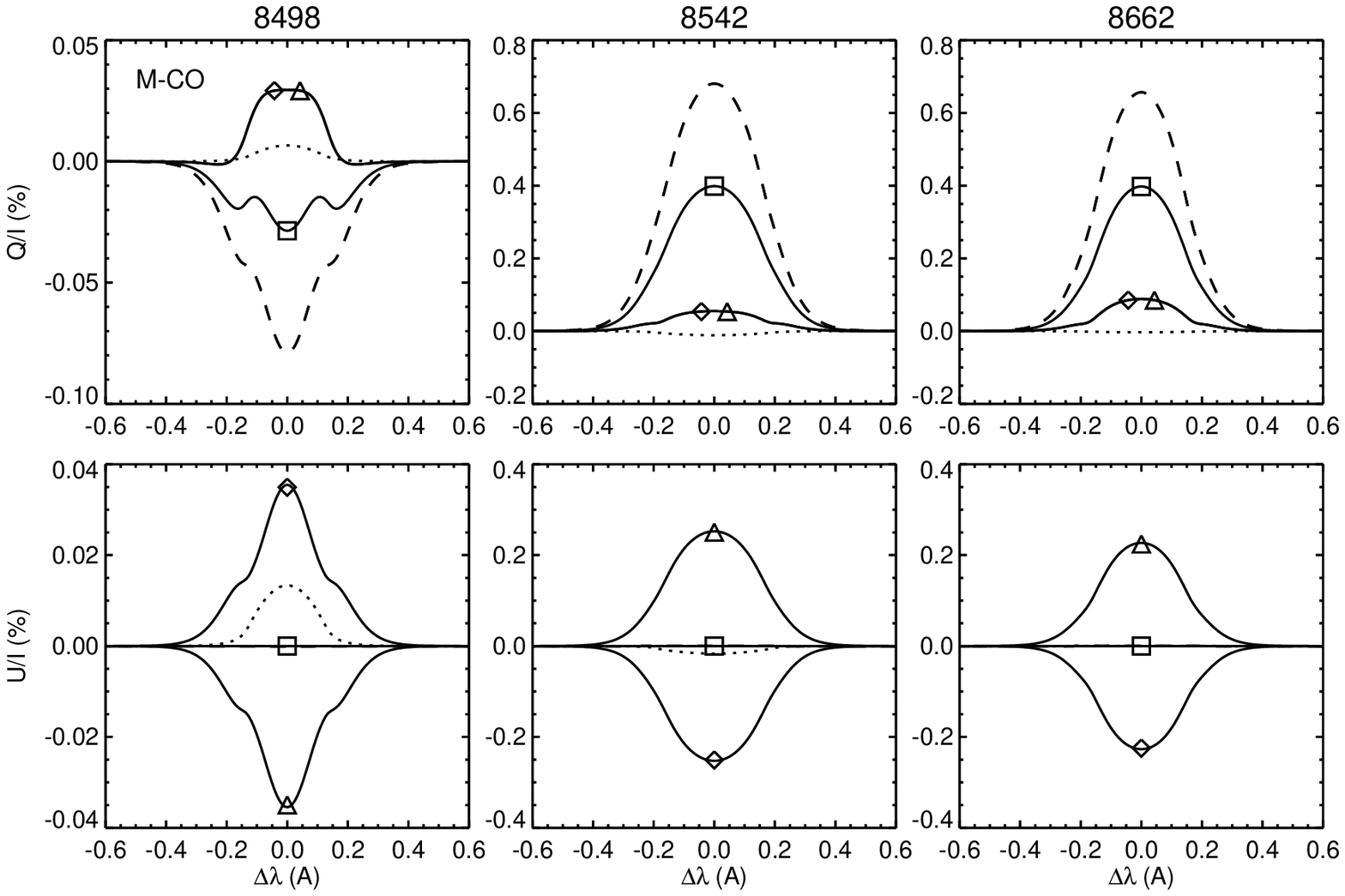}
\caption{The emergent $Q/I$ and $U/I$ profiles of the Ca {\sc ii} IR triplet calculated for a line of sight with $\mu=0.1$ in the ``hot" FAL-C model (top panels) and in the ``cool" M-CO model (bottom panels), assuming a magnetic field with the following magnetic configurations: pure scattering case (dashed lines) and the case of a magnetic field pointing to the observer with B=0.005 G (solid lines with the triangle symbol) and B=100 G (dotted lines). Actually, for the B=0.005 G case three different orientations 
of the line-of-sight are shown: along the field (triangles), perpendicularly to the field (squares), and away from the field (diamonds). The positive $Q$-direction is the parallel to the nearest limb. 
\label{fig10}}
\end{figure*}

\begin{figure*}
\plotone{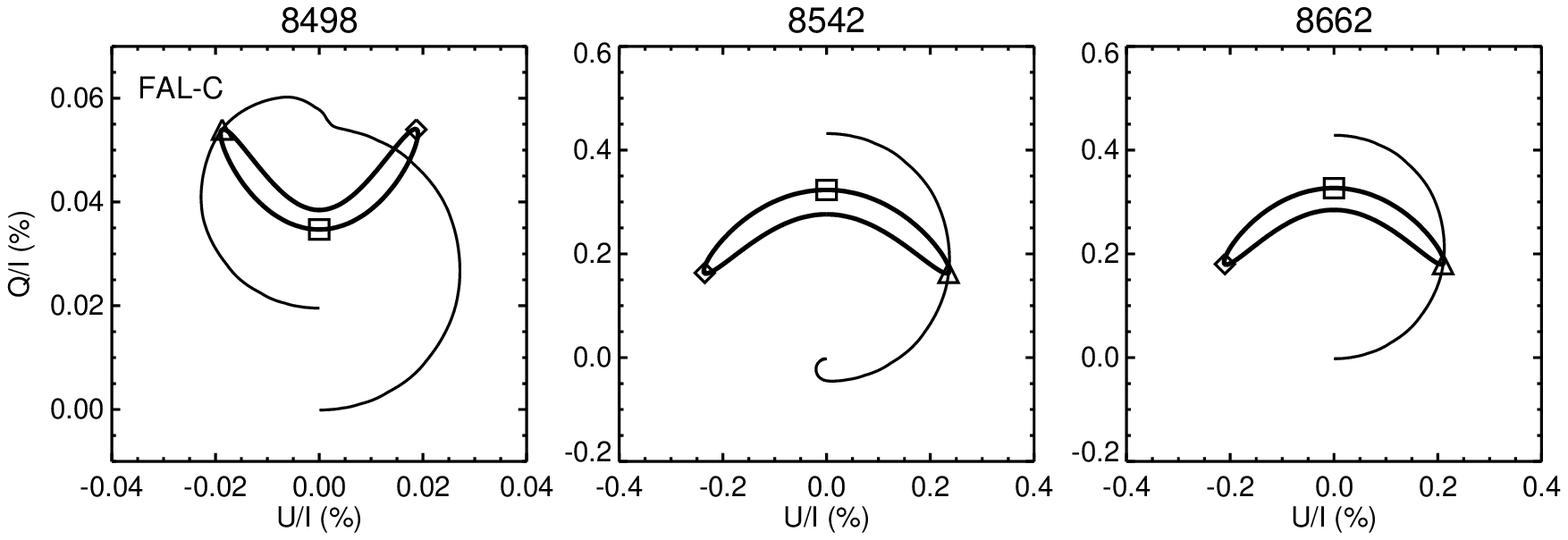}
\plotone{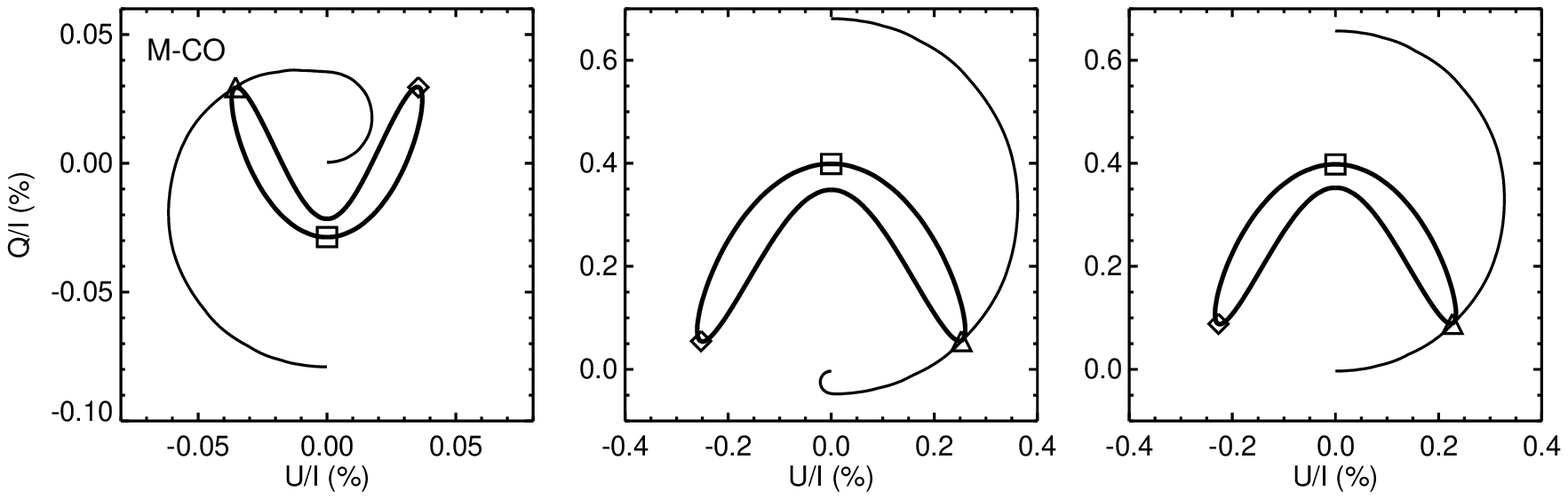}
\caption{The evolution of the line center signal in the Q/I - U/I space, for the FAL-C model (upper panels) and the M-CO model (bottom panels). Thin lines: horizontal 
magnetic field with variable strength (from B=0 G to B${\rightarrow}{\infty}$). Thick lines: 
B=0.005 G horizontal field observed from diffeerent angles. The symbols 
correspond to the particular configurations shown in Fig. 10. 
\label{fig11}}
\end{figure*}

\begin{figure}
\plotone{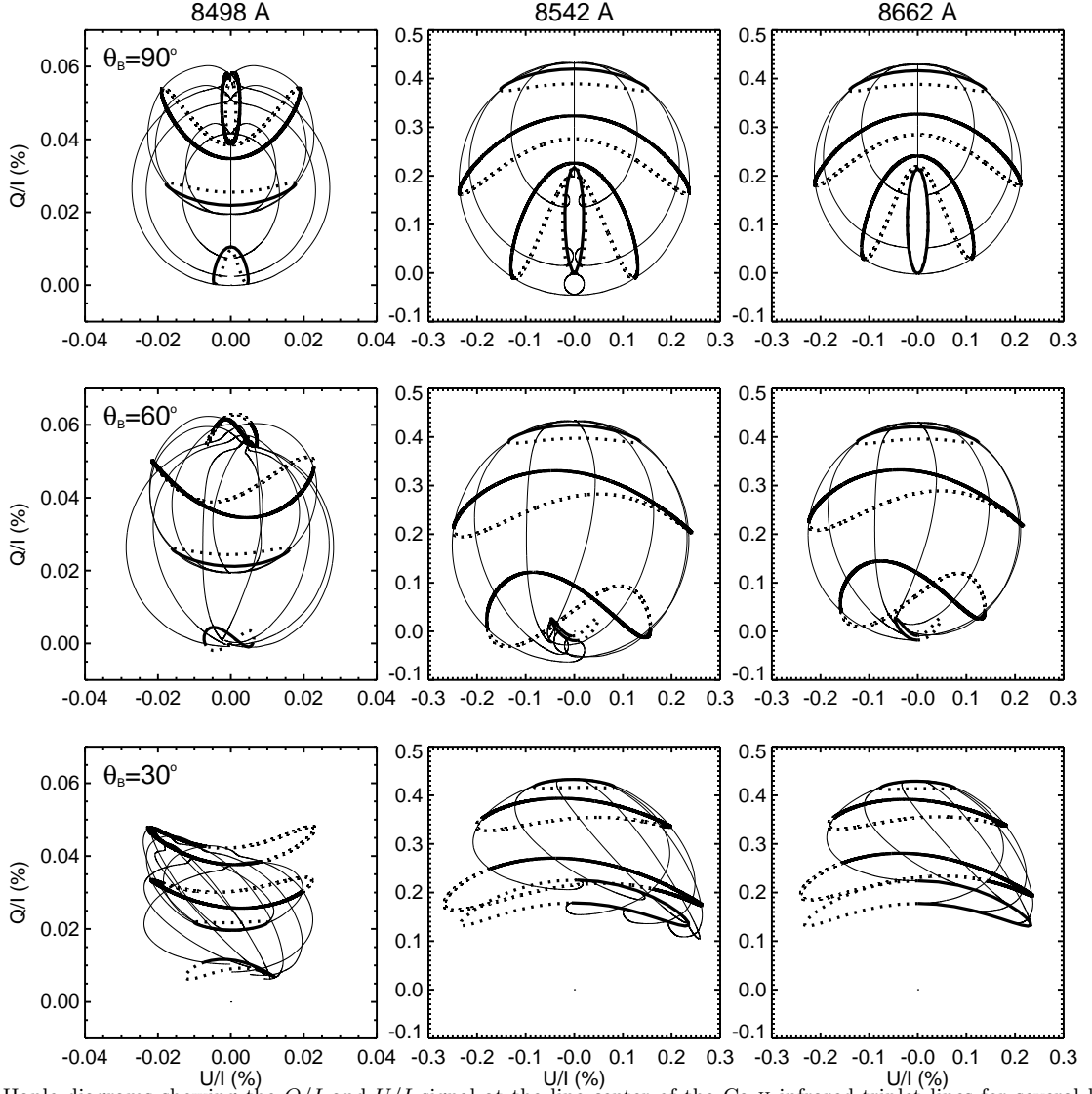}
\caption{Hanle diagrams showing the $Q/I$ and $U/I$ signal at the line center of the
Ca~{\sc ii} infrared triplet lines for several line-of-sight and magnetic 
field configurations in the FAL-C atmospheric model.
The observation is always close-to-the-limb ($\mu=0.1$).
Thick lines correspond to constant strength magnetic fields of 
$1.8\times 10^{-3}$, $5.6\times 10^{-3}$, $1.8\times 10^{-2}$, and
$18$~G when observed with $0^\circ<\chi-\chi_B<180^\circ$ (solid line)
and $180^\circ<\chi-\chi_B<360^\circ$ (dotted line),
where $\chi-\chi_B$ is the azimuth between the line-of-sight and the 
magnetic field.
(These two tracks do not coincide in the $\theta_B=90^\circ$ case because
we are not in the $90^\circ$-scattering or tangential observation limit, but 
at $\mu=0.1$).
The thin solid lines correspond to $\chi-\chi_B=0^\circ, 30^\circ, 60^\circ,
90^\circ, 120^\circ, 150^\circ$, and $180^\circ$.
Upper panels correspond to a magnetic field inclined $\theta_B=90^\circ$ with respect
to the vertical (i.e., horizontal magnetic field), central and lower panels 
correspond to inclinations of $\theta_B=60^\circ$ and $\theta_B=30^\circ$, respectively.
\label{fig12}}
\end{figure}

\begin{figure}
\plotone{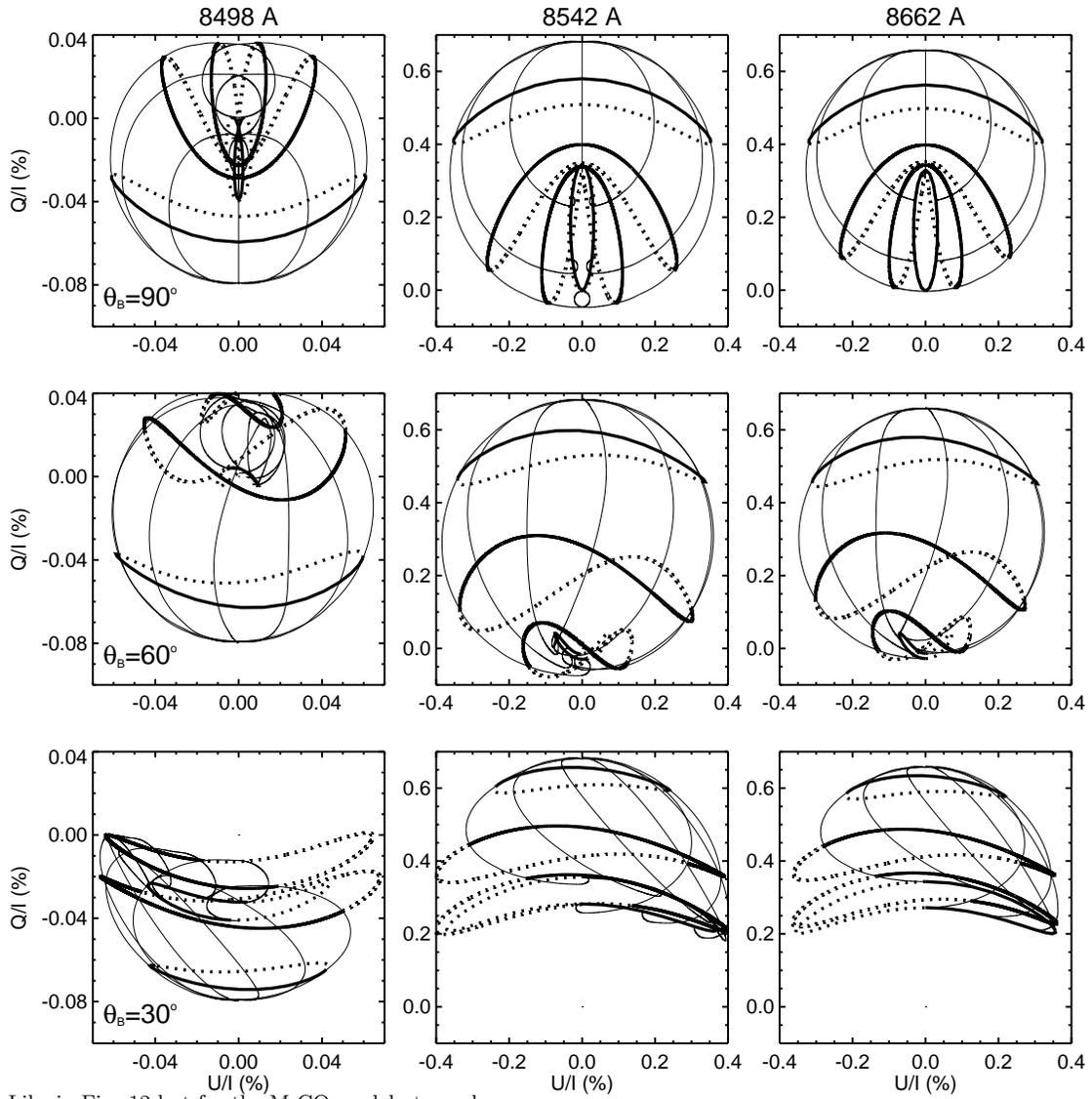}
\caption{Like in Fig. 12 but for the M-CO model atmosphere.
\label{fig13}}
\end{figure}

\begin{figure}
\plotone{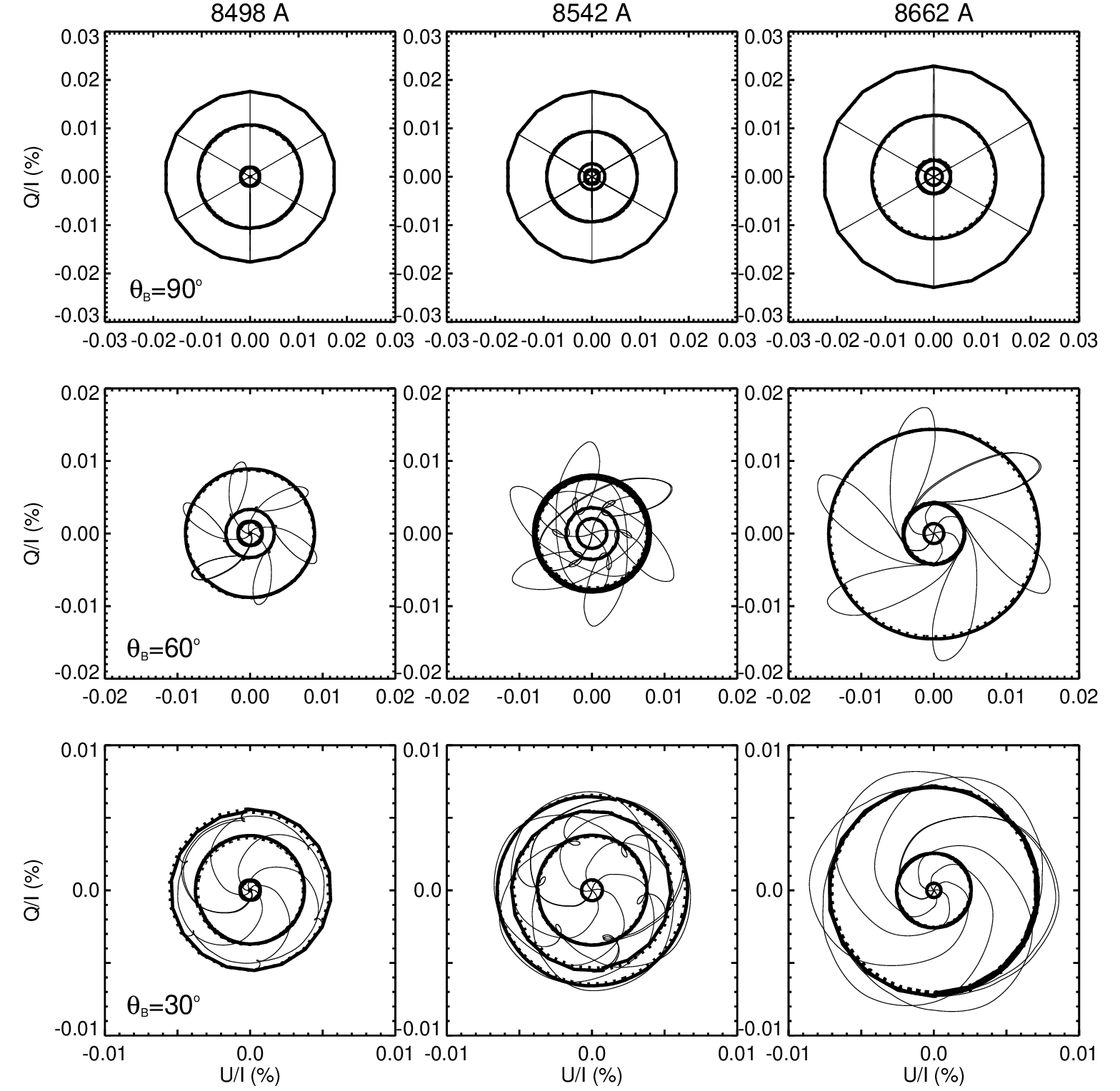}
\caption{Like in Fig. 12 but in forward scattering geometry.
\label{fig14}}
\end{figure}

\begin{figure}
\plotone{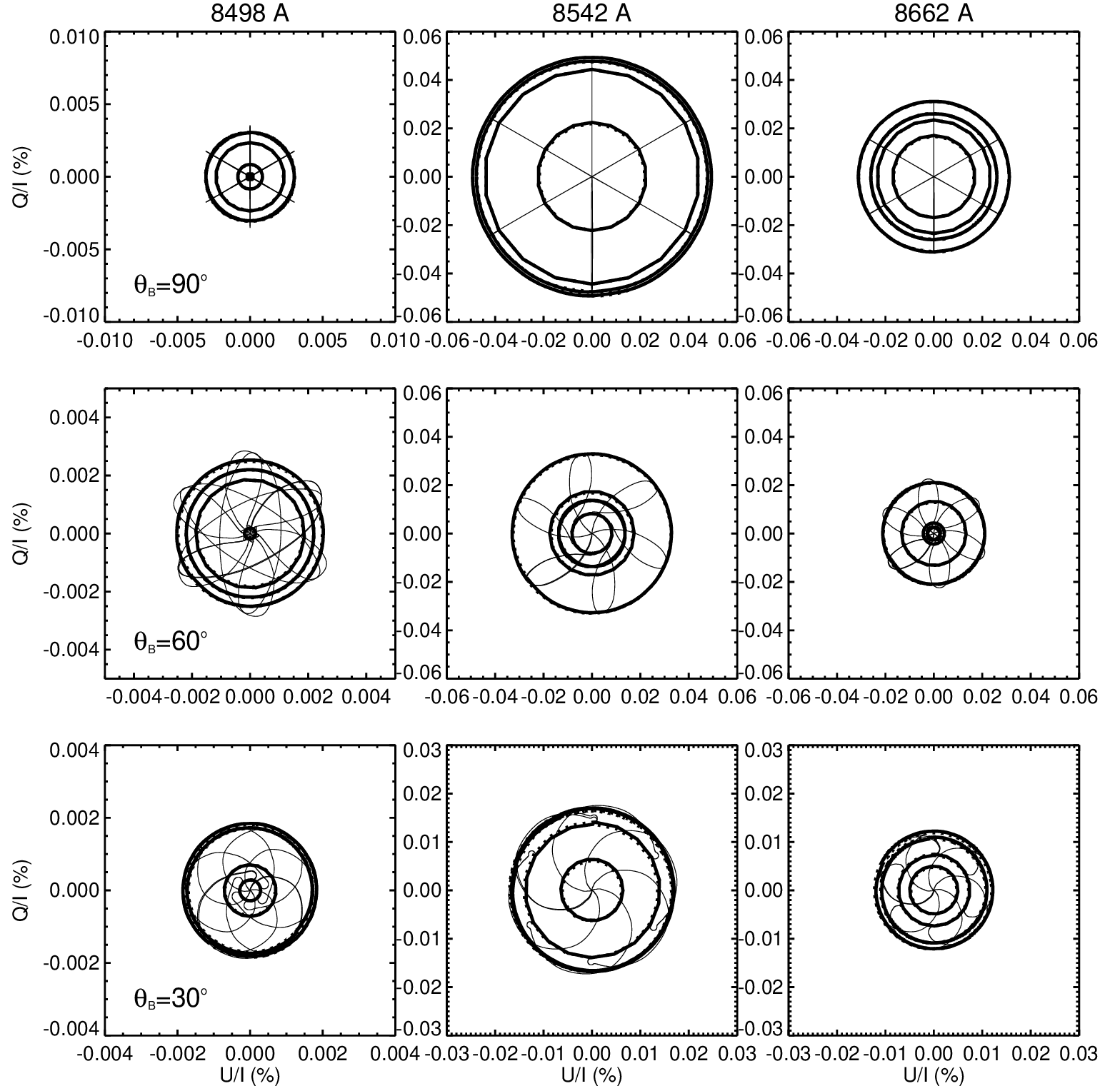}
\caption{Like in Fig. 13 but in forward scattering geometry.
\label{fig15}}
\end{figure}

\begin{figure}
\plotone{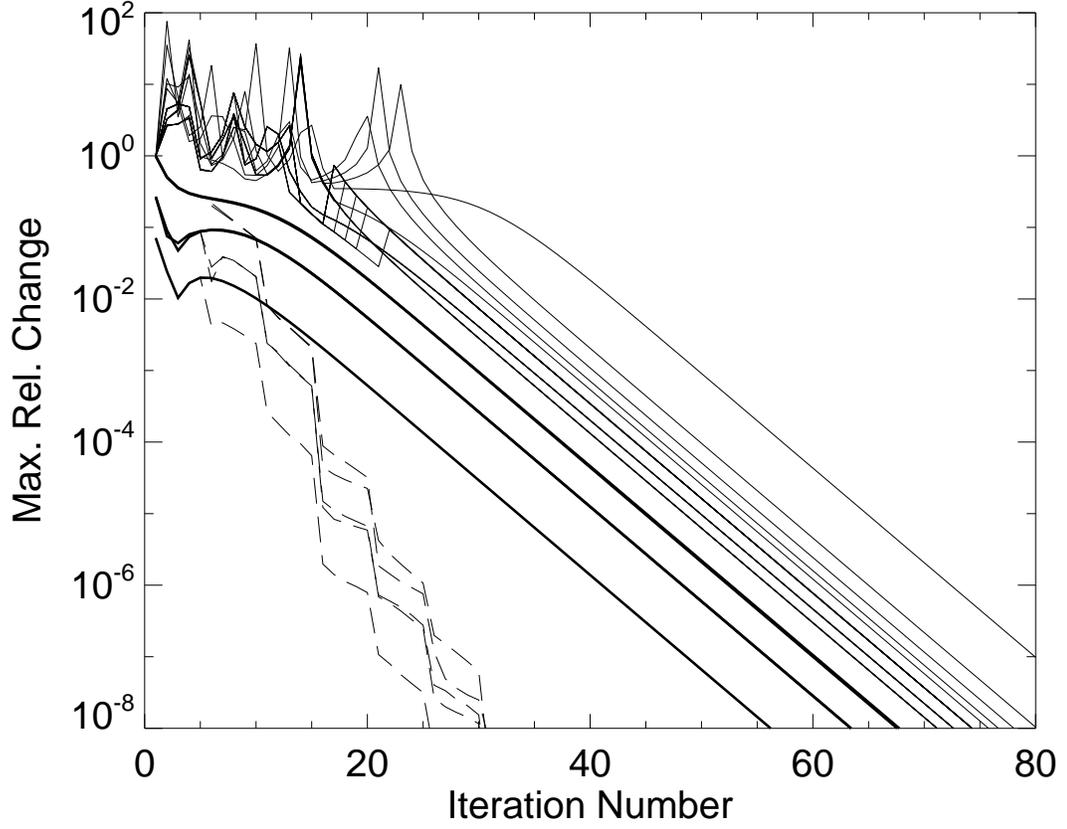}
\caption{Maximum relative change ($R_c({\rm it})={\rm Max}_{i=1, ..., N_z} 
[|\rho^K_Q(i; {\rm it}) - \rho^K_Q(i; {\rm it-1})| /|\rho^K_Q(i; {\rm it})|]$)
for each of the 29 $\rho^K_Q$ components in a radiative transfer calculation using a 5-level
atomic model of Ca {\sc ii}. Solid lines show the convergence rate of all $\rho^K_Q$ components. 
Heavy solid lines correspond to $\rho^0_0$. Note that due to superpositions
only three lines are visible while there are 5 $\rho^0_0$ components.
Dashed lines correspond to the case in which Ng-acceleration is used.
For clarity, in this case only the $\rho^0_0$ convergence rates are shown.
\label{fig16}}
\end{figure}

%%%%%%%%%%%%%%%%%%%%%%%%%%%%%%%%%%%%%%%%%%%%%%%%%%%%%%%%%%%%%%%%%
% The bibliography
%%%%%%%%%%%%%%%%%%%%%%%%%%%%%%%%%%%%%%%%%%%%%%%%%%%%%%%%%%%%%%%%%

%\bibliographystyle{apj} \bibliography{apjmnemonic,../biblio}

\end{document}